\pdfoutput=1

\documentclass[12pt,a4paper]{article}

\usepackage{ifthen}
\newboolean{pdflatex}
\setboolean{pdflatex}{true}

\newboolean{articletitles}
\setboolean{articletitles}{true}

\newboolean{uprightparticles}
\setboolean{uprightparticles}{false}

\newboolean{inbibliography}
\setboolean{inbibliography}{false}

\usepackage[top=1in, bottom=1.25in, left=1in, right=1in]{geometry}

\usepackage{microtype}
\usepackage{lineno}
\usepackage{xspace}
\usepackage{caption}

\usepackage{graphicx}
\usepackage{color}
\usepackage{colortbl}

\usepackage{amsmath}
\usepackage{amssymb}
\usepackage{amsfonts}
\usepackage{upgreek}

\newcommand*\patchAmsMathEnvironmentForLineno[1]{
\expandafter\let\csname old#1\expandafter\endcsname\csname #1\endcsname
\expandafter\let\csname oldend#1\expandafter\endcsname\csname
end#1\endcsname
 \renewenvironment{#1}
   {\linenomath\csname old#1\endcsname}
   {\csname oldend#1\endcsname\endlinenomath}
}
\newcommand*\patchBothAmsMathEnvironmentsForLineno[1]{
  \patchAmsMathEnvironmentForLineno{#1}
  \patchAmsMathEnvironmentForLineno{#1*}
}
\AtBeginDocument{
\patchBothAmsMathEnvironmentsForLineno{equation}
\patchBothAmsMathEnvironmentsForLineno{align}
\patchBothAmsMathEnvironmentsForLineno{flalign}
\patchBothAmsMathEnvironmentsForLineno{alignat}
\patchBothAmsMathEnvironmentsForLineno{gather}
\patchBothAmsMathEnvironmentsForLineno{multline}
\patchBothAmsMathEnvironmentsForLineno{eqnarray}
}

\usepackage{hyperref}
\usepackage[all]{hypcap}

\usepackage{xspace}
\usepackage{upgreek}

\def\lhcb {\mbox{LHCb}\xspace}

\def\MagUp {\mbox{\em Mag\kern -0.05em Up}\xspace}

\ifthenelse{\boolean{uprightparticles}}
{
 
 \def\Pgamma      {\ensuremath{\upgamma}\xspace}

 \def\Ppi         {\ensuremath{\uppi}\xspace}

 \def\Pphi        {\ensuremath{\upphi}\xspace}
 
 \def\Pchi        {\ensuremath{\upchi}\xspace}
 \def\Ppsi        {\ensuremath{\uppsi}\xspace}

 \def\PDelta      {\ensuremath{\Delta}\xspace}
 \def\PXi      {\ensuremath{\Xi}\xspace}
 \def\PLambda      {\ensuremath{\Lambda}\xspace}
 \def\PSigma      {\ensuremath{\Sigma}\xspace}
 \def\POmega      {\ensuremath{\Omega}\xspace}
 \def\PUpsilon      {\ensuremath{\Upsilon}\xspace}

 \def\PB      {\ensuremath{\mathrm{B}}\xspace}
 
 \def\PD      {\ensuremath{\mathrm{D}}\xspace}

 \def\PJ      {\ensuremath{\mathrm{J}}\xspace}
 \def\PK      {\ensuremath{\mathrm{K}}\xspace}

 \def\Pb      {\ensuremath{\mathrm{b}}\xspace}
 \def\Pc      {\ensuremath{\mathrm{c}}\xspace}

 \def\Pi      {\ensuremath{\mathrm{i}}\xspace}

 \def\Pp      {\ensuremath{\mathrm{p}}\xspace}

 \def\Ps      {\ensuremath{\mathrm{s}}\xspace}

}
{
 
 \def\Pgamma      {\ensuremath{\gamma}\xspace}

 \def\Ppi         {\ensuremath{\pi}\xspace}

 \def\Pphi        {\ensuremath{\phi}\xspace}
 
 \def\Pchi        {\ensuremath{\chi}\xspace}
 \def\Ppsi        {\ensuremath{\psi}\xspace}
 
 \mathchardef\PDelta="7101
 \mathchardef\PXi="7104
 \mathchardef\PLambda="7103
 \mathchardef\PSigma="7106
 \mathchardef\POmega="710A
 \mathchardef\PUpsilon="7107
 
 \def\PB      {\ensuremath{B}\xspace}
 
 \def\PD      {\ensuremath{D}\xspace}

 \def\PJ      {\ensuremath{J}\xspace}
 \def\PK      {\ensuremath{K}\xspace}

 \def\Pb      {\ensuremath{b}\xspace}
 \def\Pc      {\ensuremath{c}\xspace}

 \def\Pi      {\ensuremath{i}\xspace}

 \def\Pp      {\ensuremath{p}\xspace}

 \def\Ps      {\ensuremath{s}\xspace}

}

\makeatletter
\ifcase \@ptsize \relax
  \newcommand{\miniscule}{\@setfontsize\miniscule{4}{5}}
\or
  \newcommand{\miniscule}{\@setfontsize\miniscule{5}{6}}
\or
  \newcommand{\miniscule}{\@setfontsize\miniscule{5}{6}}
\fi
\makeatother

\DeclareRobustCommand{\optbar}[1]{\shortstack{{\miniscule (\rule[.5ex]{1.25em}{.18mm})}
  \\ [-.7ex] $#1$}}

\def\g      {{\ensuremath{\Pgamma}}\xspace}

\def\squark    {{\ensuremath{\Ps}}\xspace}

\def\cquark    {{\ensuremath{\Pc}}\xspace}

\def\bquark    {{\ensuremath{\Pb}}\xspace}

\def\pion   {{\ensuremath{\Ppi}}\xspace}

\def\pip    {{\ensuremath{\pion^+}}\xspace}
\def\pim    {{\ensuremath{\pion^-}}\xspace}

\def\kaon    {{\ensuremath{\PK}}\xspace}
  \def\Kbar    {{\kern 0.2em\overline{\kern -0.2em \PK}{}}\xspace}

\def\KorKbar    {\kern 0.18em\optbar{\kern -0.18em K}{}\xspace}

\def\Kp      {{\ensuremath{\kaon^+}}\xspace}
\def\Km      {{\ensuremath{\kaon^-}}\xspace}

\def\Kstarz  {{\ensuremath{\kaon^{*0}}}\xspace}

\def\Kstar   {{\ensuremath{\kaon^*}}\xspace}

\newcommand{\phiz}{\ensuremath{\Pphi}\xspace}

  \def\Dbar    {{\kern 0.2em\overline{\kern -0.2em \PD}{}}\xspace}
\def\D       {{\ensuremath{\PD}}\xspace}

\def\DorDbar    {\kern 0.18em\optbar{\kern -0.18em D}{}\xspace}

\def\Dstarp  {{\ensuremath{\D^{*+}}}\xspace}

\def\B       {{\ensuremath{\PB}}\xspace}
\def\Bbar    {{\ensuremath{\kern 0.18em\overline{\kern -0.18em \PB}{}}}\xspace}

\def\BorBbar    {\kern 0.18em\optbar{\kern -0.18em B}{}\xspace}
\def\Bz      {{\ensuremath{\B^0}}\xspace}
\def\Bzb     {{\ensuremath{\Bbar{}^0}}\xspace}
\def\Bu      {{\ensuremath{\B^+}}\xspace}
\def\Bub     {{\ensuremath{\B^-}}\xspace}
\def\Bp      {{\ensuremath{\Bu}}\xspace}
\def\Bm      {{\ensuremath{\Bub}}\xspace}

\def\Bd      {{\ensuremath{\B^0}}\xspace}
\def\Bs      {{\ensuremath{\B^0_\squark}}\xspace}

\def\Bdb     {{\ensuremath{\Bbar{}^0}}\xspace}

\def\jpsi     {{\ensuremath{{\PJ\mskip -3mu/\mskip -2mu\Ppsi\mskip 2mu}}}\xspace}
\def\psitwos  {{\ensuremath{\Ppsi{(2S)}}}\xspace}

\def\chicone  {{\ensuremath{\Pchi_{\cquark 1}}}\xspace}
\def\chictwo  {{\ensuremath{\Pchi_{\cquark 2}}}\xspace}
  \def\Y#1S{\ensuremath{\PUpsilon{(#1S)}}\xspace}

\def\FourS {{\Y4S}}

\def\proton      {{\ensuremath{\Pp}}\xspace}

\def\Xires       {{\ensuremath{\PXi}}\xspace}

\def\Lz          {{\ensuremath{\PLambda}}\xspace}
\def\Lbar        {{\ensuremath{\kern 0.1em\overline{\kern -0.1em\PLambda}}}\xspace}
\def\LorLbar    {\kern 0.18em\optbar{\kern -0.18em \PLambda}{}\xspace}

\def\Omegares    {{\ensuremath{\POmega}}\xspace}

\def\Lb      {{\ensuremath{\Lz^0_\bquark}}\xspace}

\def\Lc      {{\ensuremath{\Lz^+_\cquark}}\xspace}

\def\Xibm    {{\ensuremath{\Xires^-_\bquark}}\xspace}

\def\Omegab    {{\ensuremath{\Omegares^-_\bquark}}\xspace}

\def\BF         {{\ensuremath{\mathcal{B}}}\xspace}

\def\to                 {\ensuremath{\rightarrow}\xspace}

\def\AT#1     {\ensuremath{A_{\mathrm{T}}^{#1}}\xspace}

\def\C#1      {\ensuremath{\mathcal{C}_{#1}}\xspace}
\def\Cp#1     {\ensuremath{\mathcal{C}_{#1}^{'}}\xspace}
\def\Ceff#1   {\ensuremath{\mathcal{C}_{#1}^{\mathrm{(eff)}}}\xspace}
\def\Cpeff#1  {\ensuremath{\mathcal{C}_{#1}^{'\mathrm{(eff)}}}\xspace}
\def\Ope#1    {\ensuremath{\mathcal{O}_{#1}}\xspace}
\def\Opep#1   {\ensuremath{\mathcal{O}_{#1}^{'}}\xspace}

\newcommand{\tev}{\ifthenelse{\boolean{inbibliography}}{\ensuremath{~T\kern -0.05em eV}\xspace}{\ensuremath{\mathrm{\,Te\kern -0.1em V}}}\xspace}
\newcommand{\gev}{\ensuremath{\mathrm{\,Ge\kern -0.1em V}}\xspace}
\newcommand{\mev}{\ensuremath{\mathrm{\,Me\kern -0.1em V}}\xspace}
\newcommand{\kev}{\ensuremath{\mathrm{\,ke\kern -0.1em V}}\xspace}
\newcommand{\ev}{\ensuremath{\mathrm{\,e\kern -0.1em V}}\xspace}
\newcommand{\gevc}{\ensuremath{{\mathrm{\,Ge\kern -0.1em V\!/}c}}\xspace}
\newcommand{\mevc}{\ensuremath{{\mathrm{\,Me\kern -0.1em V\!/}c}}\xspace}
\newcommand{\gevcc}{\ensuremath{{\mathrm{\,Ge\kern -0.1em V\!/}c^2}}\xspace}
\newcommand{\gevgevcccc}{\ensuremath{{\mathrm{\,Ge\kern -0.1em V^2\!/}c^4}}\xspace}
\newcommand{\mevcc}{\ensuremath{{\mathrm{\,Me\kern -0.1em V\!/}c^2}}\xspace}

\def\mum  {\ensuremath{{\,\upmu\mathrm{m}}}\xspace}

\def\invfb   {\ensuremath{\mbox{\,fb}^{-1}}\xspace}

\newcommand{\chisqip}{\ensuremath{\chi^2_{\text{IP}}}\xspace}

\def\gsim{{~\raise.15em\hbox{$>$}\kern-.85em
          \lower.35em\hbox{$\sim$}~}\xspace}
\def\lsim{{~\raise.15em\hbox{$<$}\kern-.85em
          \lower.35em\hbox{$\sim$}~}\xspace}

\def\sPlot{\mbox{\em sPlot}\xspace}

\def\ptot       {\mbox{$p$}\xspace}
\def\pt         {\mbox{$p_{\mathrm{ T}}$}\xspace}

\def\evtgen     {\mbox{\textsc{EvtGen}}\xspace}

\def\geant      {\mbox{\textsc{Geant4}}\xspace}

\def\photos     {\mbox{\textsc{Photos}}\xspace}

\def\pythia     {\mbox{\textsc{Pythia}}\xspace}

\def\tell1  {TELL1\xspace}
\def\ukl1   {UKL1\xspace}

\usepackage{cite}
\usepackage{mciteplus}
\usepackage{longtable}

\begin{document}

\renewcommand{\thefootnote}{\fnsymbol{footnote}}
\setcounter{footnote}{1}

\begin{titlepage}
\pagenumbering{roman}

\vspace*{-1.5cm}
\centerline{\large EUROPEAN ORGANIZATION FOR NUCLEAR RESEARCH (CERN)}
\vspace*{1.5cm}
\noindent
\begin{tabular*}{\linewidth}{lc@{\extracolsep{\fill}}r@{\extracolsep{0pt}}}
\vspace*{-2.7cm}\mbox{\!\!\!\includegraphics[width=.14\textwidth]{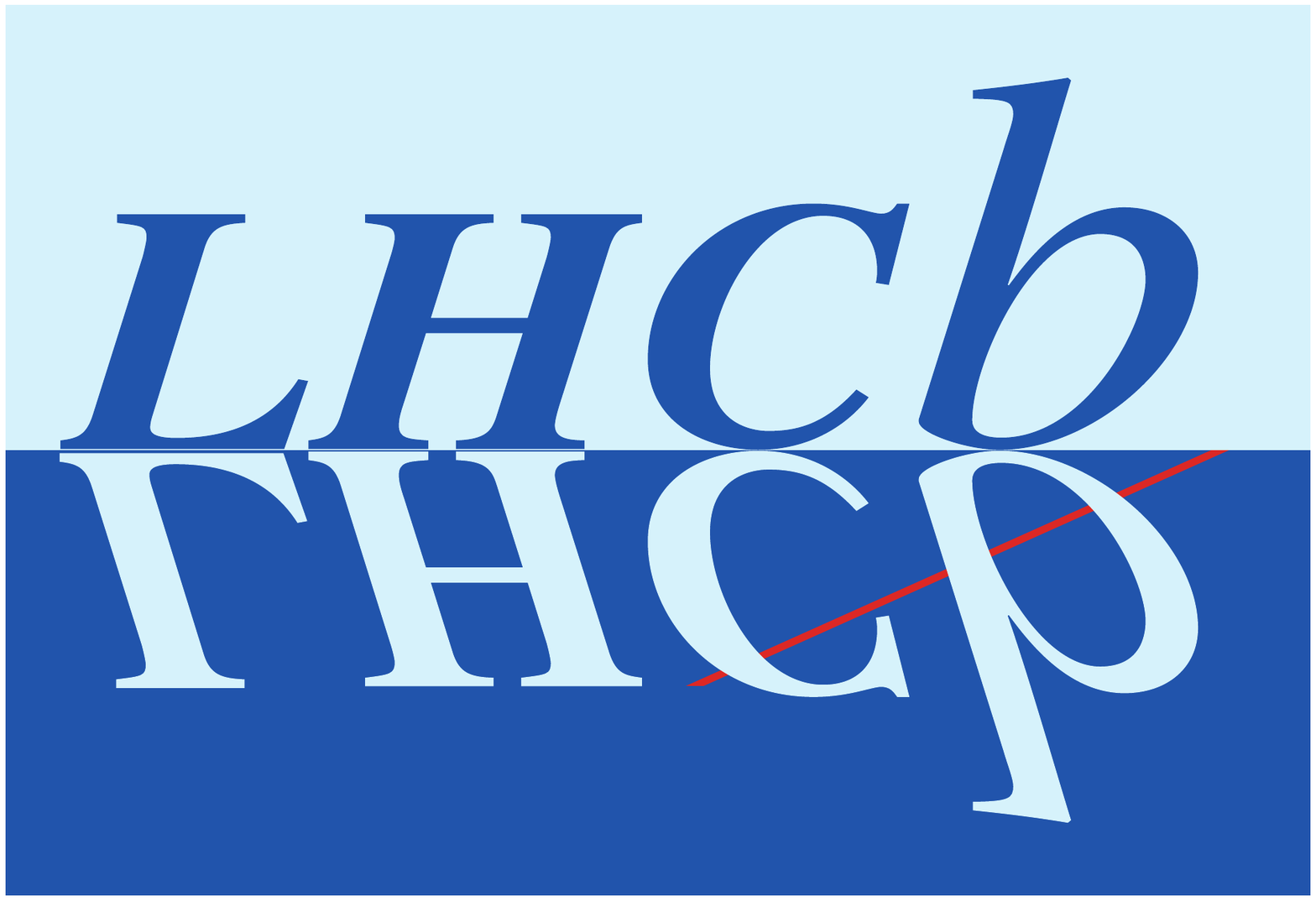}} & &
\\
 & & CERN-EP-2017-073 \\
 & & LHCb-PAPER-2017-011 \\
 & & August 9, 2017 \\
 & & \\
\end{tabular*}

\vspace*{2.0cm}

{\normalfont\bfseries\boldmath\huge
\begin{center}
  Observation of the decays $\Lb\to \chi_{c1}\proton\Km$ and $\Lb\to \chi_{c2}\proton\Km$
  \end{center}
}

\vspace*{1cm}

\begin{center}
The LHCb collaboration\footnote{Authors are listed at the end of this paper.}
\end{center}

\vspace{\fill}

\begin{abstract}
  \noindent
  The first observation of the decays $\Lb\to\chi_{c1}\proton\Km$ and
  $\Lb\to\chi_{c2}\proton\Km$
  is reported
  using a data sample corresponding to an integrated luminosity of $3.0\invfb$,
  collected by the LHCb experiment in $pp$ collisions at centre-of-mass energies of 7 and 8 TeV.
  The following ratios of branching fractions are measured
  \begin{eqnarray*}
        \dfrac{\BF(\Lb\to\chicone\proton\Km)}{\BF(\Lb\to\jpsi\proton\Km)} = 0.242 \pm 0.014 \pm 0.013 \pm 0.009\,,\\
        \dfrac{\BF(\Lb\to\chictwo\proton\Km)}{\BF(\Lb\to\jpsi\proton\Km)} = 0.248 \pm 0.020 \pm 0.014 \pm 0.009\,,\\
        \dfrac{\BF(\Lb\to\chictwo\proton\Km)}{\BF(\Lb\to\chicone\proton\Km)} = 1.02\phantom{0} \pm 0.10\phantom{0} \pm 0.02\phantom{0} \pm 0.05\,,\phantom{0}\\
  \end{eqnarray*}
  where the first uncertainty is statistical, the second systematic and the third
  due to the uncertainty on the branching fractions of the $\chicone\to\jpsi\g$ and $\chictwo\to\jpsi\g$ decays.
  Using both decay modes, the mass of the \Lb baryon is also measured to be
  \begin{center}
	  $m_\Lb = 5619.44 \pm 0.28  \pm 0.26 \mevcc\,,$
  \end{center}
  where the first and second uncertainties are statistical and systematic, respectively.
\end{abstract}

\vspace*{1.0cm}

\begin{center}
    Published as Phys.~Rev.~Lett. 119, 062001 (2017)
\end{center}

\vspace{\fill}

{\footnotesize
\centerline{\copyright~CERN on behalf of the \lhcb collaboration, licence \href{http://creativecommons.org/licenses/by/4.0/}{CC-BY-4.0}.}}
\vspace*{2mm}

\end{titlepage}

\newpage
\setcounter{page}{2}
\mbox{~}

\cleardoublepage

\renewcommand{\thefootnote}{\arabic{footnote}}
\setcounter{footnote}{0}

\pagestyle{plain}
\setcounter{page}{1}
\pagenumbering{arabic}

\begin{figure}[b]
    \centering
    \includegraphics[width=0.6\linewidth]{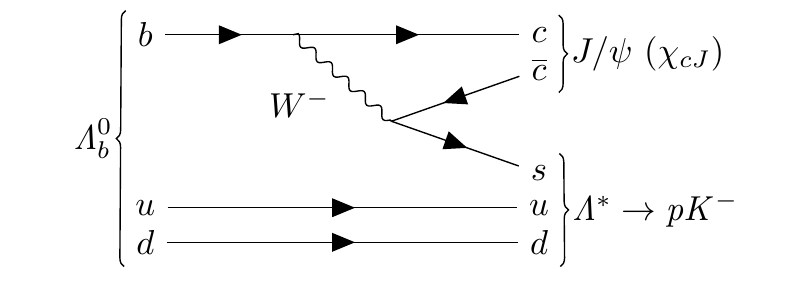}
        \caption{\small Feynman diagram of $\Lb\to\jpsi\Lz^*$ and $\Lb\to\chi_{cJ}\Lz^*$ decays,
	where $\Lz^*$ refers to an excited $\Lz$ baryon.
	}
    \label{fig:Lb2chicpKdecay}
\end{figure}

Since the birth of the quark model, it has been
speculated that hadrons could be formed from
multiquark states beyond the well-studied
quark-antiquark (meson) and three-quark (baryon)
combinations~\cite{GellMann:1964nj,Zweig:1981pd,Zweig:1964jf}.
Using a six-dimensional amplitude analysis of the $\Lb\to\jpsi\proton\Km$
decay mode, the LHCb collaboration observed the $P_c(4380)^+$ and $P_c(4450)^+$
states~\cite{LHCb-PAPER-2015-029,LHCb-PAPER-2016-009},
which are consistent with $uudc\overline{c}$ hidden-charm pentaquarks decaying to $\jpsi\proton$.
Many phenomenological models describing the dynamics of these states have been proposed, including
meson-baryon molecules~\cite{Karliner:2015ina,Chen:2015moa,Roca:2015dva},
compact pentaquarks~\cite{Maiani:2015vwa,Lebed:2015tna,Li:2015gta} and kinematic rescattering
effects~\cite{Guo:2015umn,Mikhasenko:2015vca,Liu:2015fea,Meissner:2015mza,Guo:2016bkl}.
In particular, the authors of Ref.~\cite{Guo:2015umn} noted the closeness of the
$P_c(4450)^+$ mass to the $\chi_{c1}p$ threshold and
proposed that, if the $P_c(4450)^+$ state is a rescattering effect, then it would not appear
as an enhancement near the $\chi_{c1}p$ threshold in the $\Lb \to \chi_{c1}p\Km$ decay mode,
an approach recently challenged in Ref.~\cite{Bayar:2016ftu}.
This Letter presents an initial stage in the investigation of this hypothesis by making the
first observation of $\Lb\to\chicone\proton\Km$ and $\Lb\to\chictwo\proton\Km$ decays and
measurements of their branching fractions relative to the \mbox{$\Lb\to\jpsi\proton\Km$} decay.
Throughout this Letter, the inclusion of charge-conjugated processes is implied
and the symbol $\chi_{cJ}$ is used to denote the $\chicone$ and $\chictwo$ states collectively.
All $\Lb$ decay modes considered here
proceed via the same quark-level process, whose dominant contribution is shown in Fig.~\ref{fig:Lb2chicpKdecay}.
A measurement of the \Lb baryon mass is also presented.

Previous measurements of the branching fractions
of $\B\to\chi_{cJ}K$ decays~\cite{Mizuk:2008me,Aubert:2008ae,LHCb-PAPER-2013-024}
have shown that the $\chi_{c2}$ mode is suppressed relative to the $\chi_{c1}$ mode,
in agreement with the predictions from
the factorisation approach~\cite{Beneke:2008pi}, although the suppression appears to be
lessened when additional particles are present in the final state~\cite{Bhardwaj:2015rju}.
Studying the production of $\chi_{cJ}$ mesons
in \Lb baryon decays will help to further test the factorisation approach,
as the additional spectator quark
in the baryon decay may play an important role in modifying final-state interactions.

The measurements described in this Letter are based on a data sample corresponding to 1.0\invfb
of integrated luminosity collected by the LHCb experiment in $pp$ collisions at a
centre-of-mass energy of 7\tev in 2011, and 2.0\invfb at 8\tev in 2012.
The \lhcb detector~\cite{Alves:2008zz,LHCb-DP-2014-002} is a single-arm forward
spectrometer covering the \mbox{pseudorapidity} range $2<\eta <5$,
designed for the study of particles containing \bquark or \cquark
quarks. The detector includes a high-precision tracking system
consisting of a silicon-strip vertex detector surrounding the $pp$
interaction region, a large-area silicon-strip detector located
upstream of a dipole magnet with a bending power of about
$4{\mathrm{\,Tm}}$, and three stations of silicon-strip detectors and straw
drift tubes placed downstream of the magnet.
The tracking system provides a measurement of momentum, \ptot, of charged particles with
a relative uncertainty that varies from $0.5\,\%$ at low momentum to $1.0\,\%$ at 200\gevc.
The minimum distance of a track to a primary vertex (PV), the impact parameter (IP),
is measured with a resolution of $(15+29/\pt)\mum$,
where \pt is the component of the momentum transverse to the beam, in\,\gevc.
Different types of charged hadrons are distinguished using information
from two ring-imaging Cherenkov detectors.
Photons, electrons and hadrons are identified by a calorimeter system consisting of
scintillating-pad and preshower detectors, an electromagnetic
calorimeter and a hadronic calorimeter. Muons are identified by a
system composed of alternating layers of iron and multiwire
proportional chambers.
The online event selection is performed by a trigger,
which consists of a hardware stage, based on information from the calorimeter and muon
systems, followed by a software stage, which applies a full event
reconstruction.
The software trigger selects events that contain a pair of oppositely charged muons
that form a vertex that is significantly separated from all PVs.

In the simulation, $pp$ collisions are generated using \pythia8~\cite{Sjostrand:2006za,*Sjostrand:2007gs}
with a specific LHCb configuration~\cite{LHCb-PROC-2010-056}.
Decays of hadronic particles are described by \evtgen~\cite{Lange:2001uf},
in which final-state radiation is generated using \photos~\cite{Golonka:2005pn}.
The interaction of the generated particles with the detector, and its response, are implemented using the \geant
toolkit~\cite{Allison:2006ve, *Agostinelli:2002hh} as described in
Ref.~\cite{LHCb-PROC-2011-006}. The products of the $\Lb$ decays are generated uniformly within the available phase space.

The $\Lb\to\chi_{cJ}p\Km$ and $\Lb\to\jpsi p\Km$ candidates are reconstructed via the decays
$\chi_{cJ} \to \jpsi\gamma$ and $\jpsi\to\mu^+\mu^-$. To separate signal from background,
an offline selection is applied after the trigger,
consisting of a loose preselection followed by a multivariate classifier based on a
gradient-boosted decision tree (GBDT)~\cite{Breiman}.

The $\jpsi$ candidates are formed from two oppositely charged particles with
$\pt > 550 \mevc$, identified as muons and consistent with originating from a common vertex, but
inconsistent with originating from any PV.
The invariant mass of the $\mu^+\mu^-$ pair is required to be in the range $[3000, 3170] \mevcc$.
The $\chi_{cJ}$ candidates are formed from a \jpsi candidate and a photon with
$\pt > 700 \mevc$.
Photons that are consistent with originating from a $\pi^0$ meson when combined with
any other photon in the event are removed.
The invariant mass of the $\mu^+\mu^-\gamma$ combination is required to be in the range
$[3400, 3700] \mevcc$. In the following, the notation $[c\overline{c}]$ will be used to refer to
the initial $\jpsi$ or $\chi_{cJ}$ candidate from the \Lb baryon decay,
while the notation $m(\jpsi X)$ or $m(\chi_{cJ} X)$ denotes an invariant mass
that has been calculated with a mass constraint applied to the $\jpsi$ or $\chi_{cJ}$ candidate.

The \Lb candidates are formed from a $[c\overline{c}]$ candidate and two good-quality oppositely charged
tracks each with $\pt > 200 \mevc$, identified as a proton and kaon.
Both tracks are required to be significantly displaced from any PV.
A kinematic fit~\cite{Hulsbergen:2005pu} is applied to the \Lb candidate, with the \jpsi and $\chi_{c1}$
masses constrained to their known values~\cite{PDG2016},
and the \Lb candidate constrained to point back to a PV.
This has the effect of producing separated peaks for the two decay modes.
The mass resolution for $\Lb\to\chi_{c2}p\Km$ decays is lower compared to that
for $\Lb\to\chi_{c1}p\Km$ decays due to the wrong mass hypothesis of the $[c\overline{c}]$ candidate.

Contributions from $\Bd\to[c\overline{c}] \Kp\pi^-$  ($\Bs\to[c\overline{c}]\phi$, $\phi\to\Kp\Km$) decays,
where the $\pi^-$ ($\Km$) is misidentified as an antiproton, are suppressed by placing tighter
particle identification
requirements on the misidentified hadron for candidates with an invariant
mass within 30\mevcc of the \Bd (\Bs) mass~\cite{PDG2016}
when evaluated using $\pim$ or $\Km$ mass assignments for the antiproton candidates.
Typical mis-identification probabilities are $5-15$\,\%, dependent on the particle
momentum~\cite{LHCb-DP-2012-003}.
In the $\Lb\to\chi_{cJ}p\Km$ samples, small contributions
from $\Bd\to\jpsi \Kp\pi^-$ or $\Bs\to\jpsi\phi$ decays, where in addition to the misidentification
the \jpsi meson is combined with a random photon in the event,
are removed  by the requirement that $m(\jpsi\Kp\pim)$ or $m(\jpsi\Kp\Km)$ are within
30\mevcc of the \Bd or \Bs mass.
Additionally, $\phi\to\Kp\Km$ decays are vetoed by removing all candidates
where the invariant mass of the $p\Km$ combination is within 12\mevcc of the known $\phi$
meson mass when the kaon mass is used instead of the proton mass.
Further misreconstructed backgrounds are studied using a
fast simulation package~\cite{Cowan:2016tnm} and are found to have no peaking components
in the invariant mass window of interest.

The GBDT is used to further suppress the combinatorial background.
It is trained on a simulated sample of $\Lb\to\chi_{c1}p\Km$ decays
for the signal and candidates from data with $m(\chi_{c1}p\Km)$ in the range
$[5700,5800] \mevcc$ for the background.
Twelve variables are used as input.
The first of these is the $\chi^2$ value obtained from a kinematic fit with the $\Lb$ candidate constrained to point back to a PV and a mass constraint applied to the $\jpsi$. In addition, for the signal mode, $\chicone$ or $\chictwo$ mass constraints are applied, with the smaller $\chi^2$ values being used;
note that this differs from the fit used to separate the mass peaks,
which does not include a $\chictwo$ constraint. The remaining variables
are the $\pt$ of the $\Lb$, proton
and kaon; the $\Lb$ decay-length significance; the cosine of the angle between the momentum of the $\Lb$ candidate and its displacement from the PV;
the proton and kaon $\chisqip$, defined as the difference in the vertex-fit $\chi^2$
of the PV when reconstructed with and without the considered particle;
and the estimated probabilities that the two muons, kaon and proton are correctly identified by the particle
identification detectors.

Prior to the training, several modifications
are made to the simulated samples to better match the kinematic
distributions observed in data. First, the simulated $\Lb\to\jpsi p\Km$ events are weighted
according to the six-dimensional amplitude model developed in Ref.~\cite{LHCb-PAPER-2015-029}.
Second, a multidimensional gradient-boosting algorithm~\cite{Rogozhnikov:2016bdp}
is used to weight the simulated $\Lb\to\jpsi p\Km$ decays such that the distributions of
\Lb pseudorapidity, the number of tracks in the event and the GBDT training variables (apart from those
related to particle identification) match those observed in the
preselected background-subtracted \mbox{$\Lb\to\jpsi p\Km$} data sample. These weights are also
applied to the simulated \mbox{$\Lb\to\chi_{cJ} p\Km$} samples.
Finally, the simulated distributions of
the particle identification variables for the muon, proton and kaon candidates are
resampled from data calibration samples ($\Dstarp\to\D\pip$, $\jpsi\to\mu^+\mu^-$, $\Lz\to p\pi^-$
and $\Lc\to p\Km\pip$ decays) in bins of track $p$, \pt and the number of tracks.

The optimal working point for the GBDT response
is chosen by maximising a figure of merit, $S/\sqrt{S+B}$, where $S=S_0\epsilon$ and $B$
are the expected signal and background yields within $\pm 20 \mevcc$ of the
known \Lb baryon mass~\cite{PDG2016},  $S_0$ is the signal yield
determined from data without any cut on the GBDT response, and $\epsilon$ is the relative efficiency of the GBDT selection,
evaluated using the simulated sample. The \Lb
mass sidebands from the data are used to estimate $B$. The same
GBDT and working point are used for the $\Lb\to\jpsi p\Km$ normalisation mode. The GBDT selection efficiencies
are $78\,\%$, $75\,\%$ and $68\,\%$ for the $\Lb\to\chi_{c1}p\Km$, $\Lb\to\chi_{c2}p\Km$ and $\Lb\to\jpsi p\Km$
channels, respectively.

After applying the GBDT requirement, $(2.9\pm0.4)\,\%$ of the selected events contain multiple
$\Lb\to\chi_{cJ} p \Km$ candidates. 
In approximately $80\,\%$ of these cases, the same $\jpsi p\Km$ combination is
combined with an additional, unrelated, photon in the event. The  results reported
in this Letter retain all candidates and the reported branching fractions are corrected
to account for this. The correction factor is $0.993\pm0.006$ ($0.986 \pm 0.009$) for
$\Lb\to\chicone p\Km$ ($\Lb\to\chictwo p\Km$) decays, which is evaluated using a combination
of the simulated samples and pseudoexperiments.
The larger width of the $\Lb\to\chi_{c2}p\Km$ component
leads to the larger uncertainty on the correction factor.
For the selected $\Lb\to\jpsi p \Km$ sample, $(0.75\pm0.05)\,\%$ of the events have multiple candidates.

\begin{figure}[t]
	\setlength{\unitlength}{\textwidth}
	\centering
	\begin{picture}(0.98,0.32)
		\put(0,0){
	\includegraphics[width=0.49\linewidth]{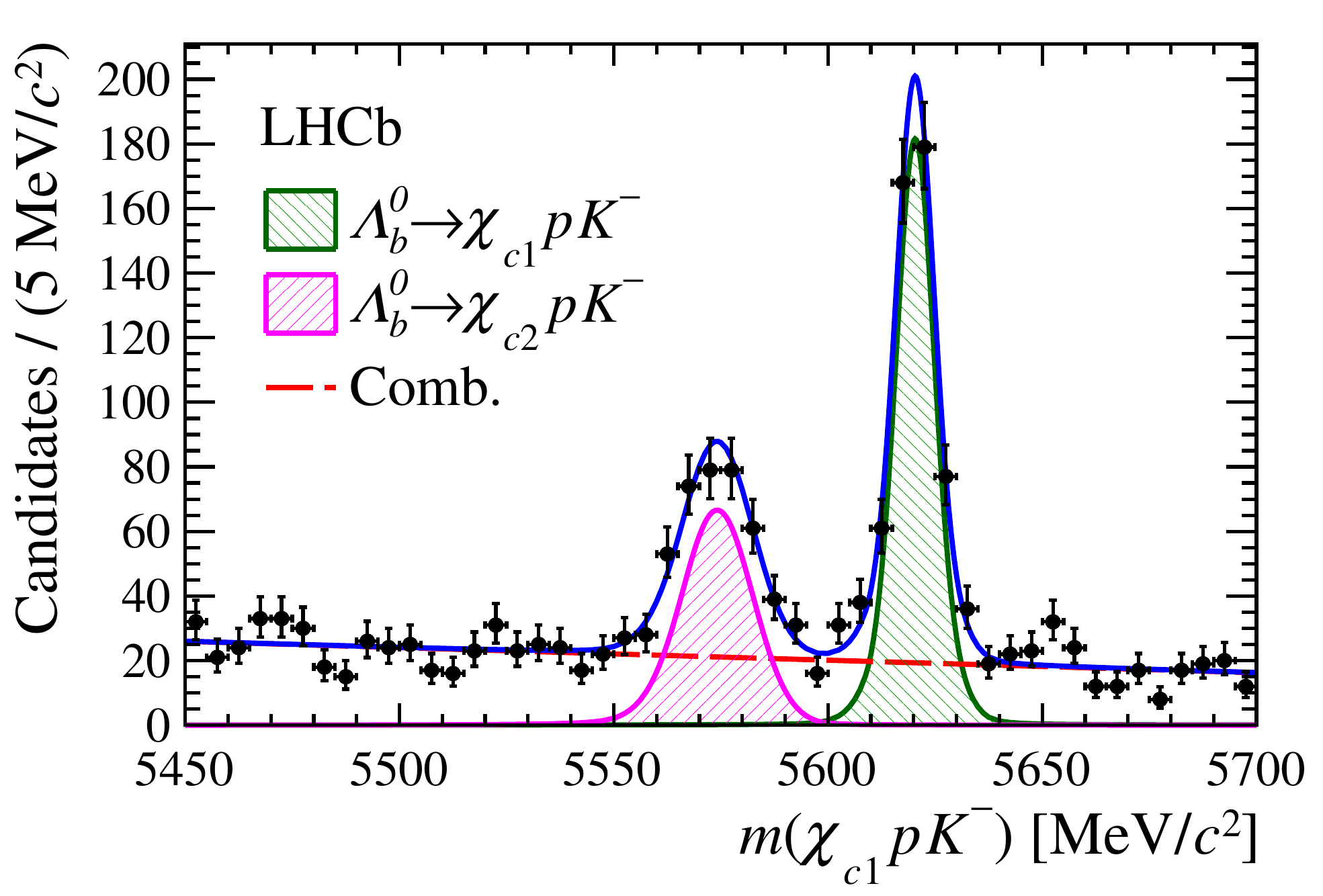}
		}
		\put(0.49,0){
	\includegraphics[width=0.49\linewidth]{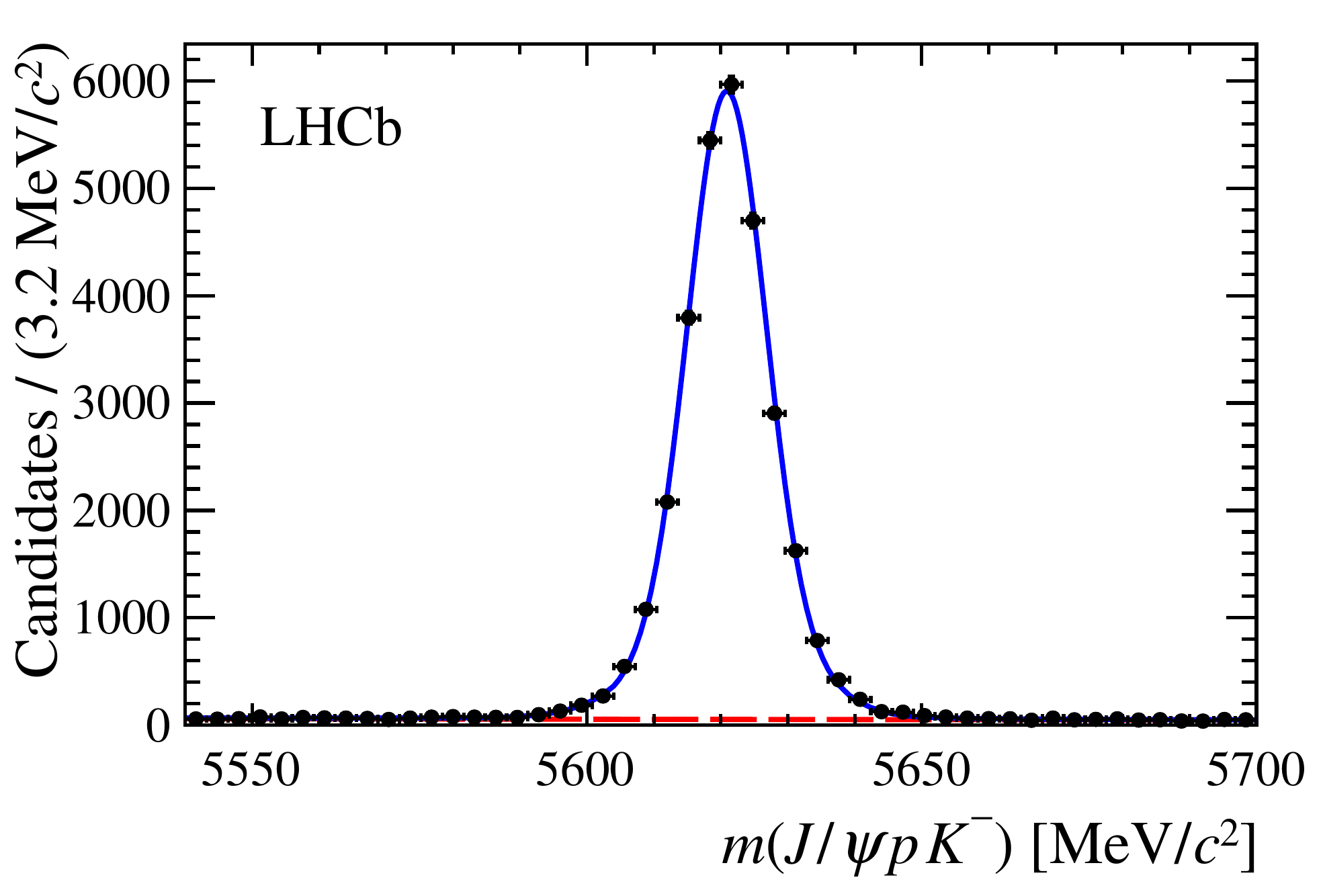}
		}
    		\put(0.42,0.28){(a)}
    		\put(0.91,0.28){(b)}
	\end{picture}
	\centering
	\caption{\small
		Fits to the (a) $\Lb\to\chicone\proton\Km$ and (b) $\Lb\to\jpsi\proton\Km$
		invariant mass distributions. Data points are shown in black
		and the results of the fits are shown as solid blue lines.
		The components are
		$\Lb\to\chicone\proton\Km$ and
		$\Lb\to\chictwo\proton\Km$ signal
		and combinatorial background (Comb.).
	}
	\label{fig:fits}
\end{figure}

Extended unbinned maximum-likelihood fits are performed to the distributions of
$m(\chi_{c1}\proton\Km)$ and $m(\jpsi\proton\Km)$ for the signal and normalisation modes, respectively.
The fit models consist of signal components, each described by the sum of two Crystal Ball (CB)
functions~\cite{Skwarnicki:1986xj}
with a common mean and power-law tails on both sides, and a linear combinatorial background component.
Due to the
small $\chi_{c0}\to\jpsi\gamma$ branching fraction~\cite{PDG2016} the contribution from the $\chi_{c0}$
mode is negligible.
Several parameters of the signal shapes are determined from fits to simulated samples.
These include the tail parameters of the CB functions, the ratio of the
widths of the two CB functions, and their relative normalisations.
The $\Lb\to\chictwo\proton\Km$ signal component is shifted to a lower mass in $m(\chi_{c1}\proton\Km)$ due to the $\chicone$ mass constraint.
The signal and background yields, the gradient of the background shape
and the mean of each signal component are free parameters in the fit to data.
In addition, the widths of the $\chi_{c1}$ and $\chi_{c2}$ components in the fit to $m(\chi_{c1}\proton\Km)$
are allowed to differ from simulation by a common scaling factor,
while the width of the narrower CB function in the $\Lb\to\jpsi\proton\Km$ signal component is a free
parameter in the fit to data.
The results of these fits are shown in Fig.~\ref{fig:fits}.
The measured yields are $453 \pm 25$, $285 \pm 23$ and $29\,815 \pm 178$ for the $\chicone$, $\chictwo$ and $\jpsi$ modes, respectively.
The significance of each of the signal components in the fit to
$m(\chi_{c1}\proton\Km)$ is calculated using Wilks' theorem~\cite{Wilks:1938dza}. This gives
statistical significances of $29$ and $17$ standard deviations for the decay modes with $\chi_{c1}$ and $\chi_{c2}$, respectively.

\begin{figure}[t]
	\setlength{\unitlength}{\textwidth}
	\centering
	\begin{picture}(0.98,0.32)
		\put(0,0){
	\includegraphics[width=0.42\linewidth]{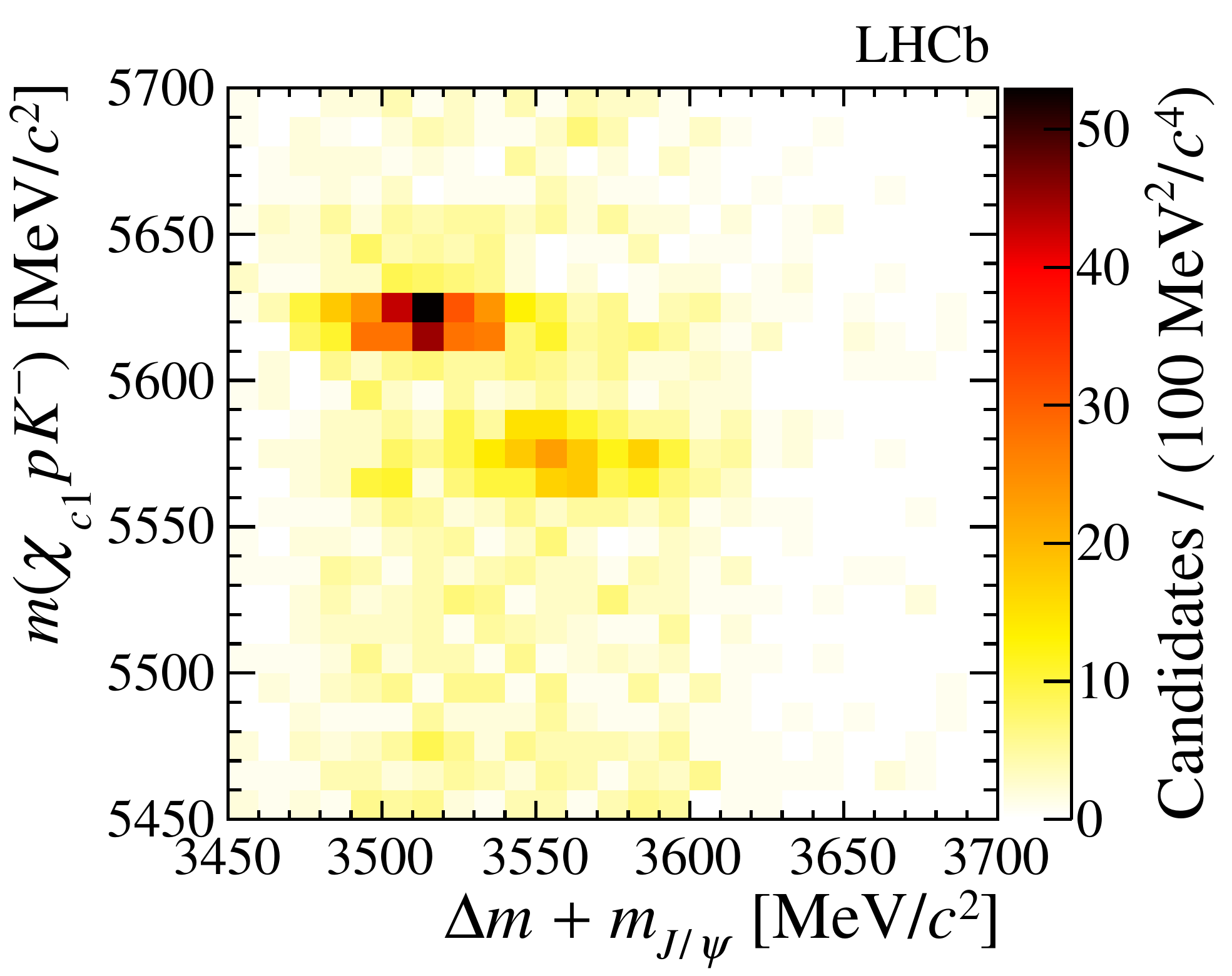}
		}
    		\put(0.30,0.27){(a)}
		\put(0.49,0){
	\includegraphics[width=0.49\linewidth]{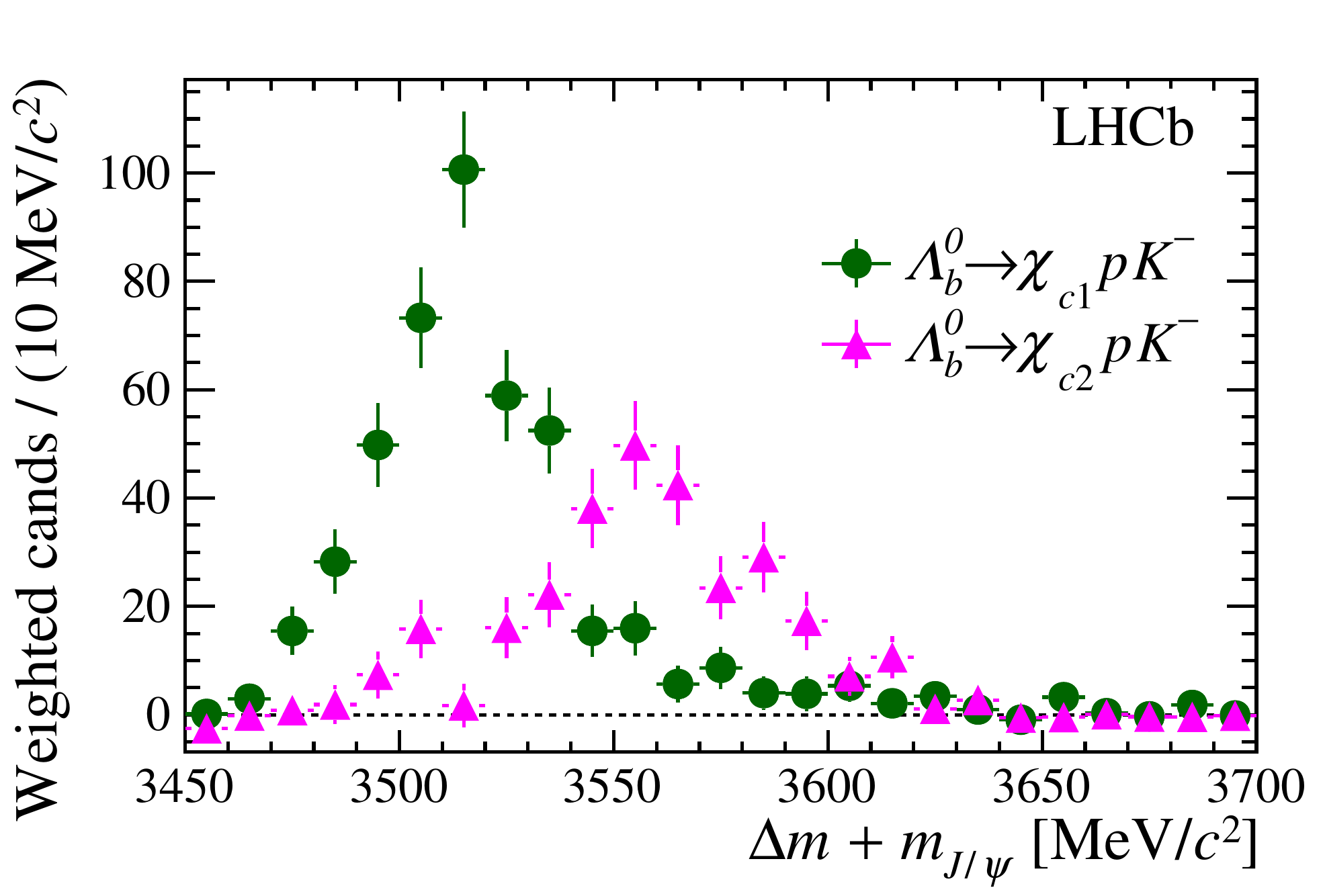}
		}
    		\put(0.58,0.27){(b)}
	\end{picture}
	\caption{\small
		Distributions of (a) $m(\chicone\proton\Km)$ versus
		$\Delta m + m_{\jpsi}$, where $\Delta m$ is $m(\mu^+\mu^-\gamma) - m(\mu^+\mu^-)$
		and $m_{\jpsi}$ is the known mass of the \jpsi meson~\cite{PDG2016}, and
		(b) $\Delta m + m_{\jpsi}$
		for background-subtracted $\Lb\to\chicone\proton\Km$ (green circles)
		and $\Lb\to\chictwo\proton\Km$ (magenta triangles) candidates (cands).
	}
	\label{fig:chic_mass}
\end{figure}

Simulated  samples are used to determine, for each decay mode, the
reconstruction and selection efficiency
as a function of the Dalitz plot coordinates~\cite{Dalitz:1953cp}, $m^2(\proton\Km)$ and $m^2([c\overline{c}]\proton)$.
This approach focuses on the dimensions where the efficiency variation is expected to be largest whilst
averaging over other dimensions in the phase space of $\Lb\to [c\overline{c}] p \Km$ decays.
This assumption is treated as a source of systematic uncertainty.
The efficiency of each mode, $\epsilon_{[c\overline{c}]}$,
is due to geometric acceptance, reconstruction and selection including the GBDT.
The ratios of the phase-space-averaged values of these efficiencies are
$0.182 \pm 0.005$ for $\epsilon_{\chicone}/\epsilon_{\jpsi}$,
and $0.196 \pm 0.005$ for $\epsilon_{\chictwo}/\epsilon_{\jpsi}$,
where the uncertainties are due to the size of the simulated samples.
This includes a correction factor for the  $\chi_{cJ}$ decay modes to account for differences in the photon reconstruction efficiency between data and
simulation~\cite{LHCb-PAPER-2013-024,Govorkova:2015vqa}.

An efficiency-corrected, background-subtracted yield is determined for each decay mode as
$N^{\rm corr}([c\overline{c}]\proton\Km)
	  = \sum_i w_i/\epsilon_{[c\overline{c}],i}$,
where the index $i$ runs over all candidates in the fit range.
The $w_i$ are weights determined using the \sPlot background-subtraction technique~\cite{Pivk:2004ty},
which project out the signal component from the combined signal plus background densities using the $[c\overline{c}]\proton\Km$ invariant mass as discriminating variable.
The corrected yields are found to be about $99\,700 \pm 5300$, $57\,800 \pm 4400$ and $1\,213\,500 \pm 7300$ for the $\chicone$, $\chictwo$ and $\jpsi$ decay modes, respectively,
where the uncertainties are determined from the sum in quadrature of the event weights.
Since these weights are determined from a fit in which all shape parameters are fixed,
following the \sPlot prescription, the effect on the yield uncertainty due to the statistical uncertainties on
these parameters is not included. To quantify this, the unweighted fit is performed with the
shape parameters fixed and then free, and the difference in quadrature of the uncertainties
is found to be 1.5\,\% of the yield for $\Lb\to\chi_{c1}\proton\Km$ decays
and 2.1\,\% for $\Lb\to\chi_{c2}\proton\Km$. A corresponding uncertainty is added in
quadrature to that on the efficiency-corrected yield. The effect is negligible for the much larger
$\Lb\to\jpsi\proton\Km$ signal.

The ratios of branching fractions are  determined as
\begin{eqnarray}\label{eqn:bfratio}
  R_{1(2)} &\equiv&
  \frac{
  {\cal B}\left(\Lb\to\chi_{c1(2)}\proton\Km\right)}{
    {\cal B}\left(\Lb\to\jpsi\proton\Km\right)} =
  \frac{
  N^{\rm corr}(\chi_{c1(2)}\proton\Km)}{
  N^{\rm corr}(\jpsi\proton\Km)} \times \frac{1}{\BF(\chi_{c1(2)}\to\jpsi\g)} \, ,
\\
  R_{2/1} &\equiv&
  \frac{
  {\cal B}\left(\Lb\to\chictwo\proton\Km\right)}{
  {\cal B}\left(\Lb\to\chicone\proton\Km\right)} =
  \frac{
  N^{\rm corr}(\chictwo\proton\Km)}{
  N^{\rm corr}(\chicone\proton\Km)} \times \frac{\BF(\chicone\to\jpsi\g)}{\BF(\chictwo\to\jpsi\g)} \, ,
\end{eqnarray}
where the branching fraction of the $\chicone\to\jpsi\gamma$ $(\chictwo\to\jpsi\gamma)$ decay is taken to be $(33.9\pm1.2)\,\%$ ($(19.2\pm0.7)\,\%$)~\cite{PDG2016}. Figure~\ref{fig:chic_mass}(a)
shows the
distribution of
$\Delta m + m_{\jpsi}$, where $\Delta m$ is $m(\mu^+\mu^-\gamma) - m(\mu^+\mu^-)$
and $m_{\jpsi}$ is the known mass of the \jpsi meson~\cite{PDG2016},
while Fig.~\ref{fig:chic_mass}(b) shows
$\Delta m + m_{\jpsi}$
for background-subtracted $\Lb\to\chi_{cJ}\proton\Km$ candidates.
Both distributions show clear enhancements at the known masses of the $\chi_{cJ}$ mesons.

The $\Lb\to\chi_{cJ}\proton\Km$ data sample is also used to make a measurement of the $\Lb$ mass, $m_{\Lb}$.
The momenta of the particles are scaled to
account for known miscalibration of the detector~\cite{LHCb-PAPER-2013-011}.
In addition, the separation between the $\Lb\to\chicone\proton\Km$ and $\Lb\to\chictwo\proton\Km$
components in the $m(\chi_{c1}\proton\Km)$ spectrum is fixed to the known mass difference between
the $\chi_{c1}$ and $\chi_{c2}$ mesons~\cite{PDG2016},
to obtain a single measurement of $m_{\Lb}$ using both decay modes.
The mass fit is repeated after these changes, yielding
$m_{\Lb}= 5619.44 \pm 0.28 \mevcc$.

Systematic uncertainties on the ratios of branching fractions are assigned due to imperfect knowledge
of the trigger efficiency and data/simulation discrepancies in the photon reconstruction
as well as for uncertainties on the corrections applied to the simulated data
(kinematic reweighting, phase-space weighting and particle identification resampling),
the treatment of multiple candidates, the limited size of simulated data samples,
and the models for the signal and background components in the fits.
The per-candidate efficiencies as a function of the Dalitz plot coordinates are also replaced
by phase-space-averaged efficiencies, and the difference with the nominal result is
assigned as a systematic uncertainty.

The systematic uncertainties due to the trigger and photon reconstruction are taken from previous LHCb studies~\cite{LHCb-PAPER-2013-024,LHCb-PAPER-2012-010,Govorkova:2015vqa}.
The uncertainties assigned due to the kinematic weighting are evaluated by repeating the analysis
with alternative efficiency histograms that make use of a simplified
three-dimensional weighting procedure
with only the \Lb pseudorapidity, \Lb \pt and event track-multiplicity.
The uncertainty on the correction for multiple candidates is assigned as a systematic
uncertainty.
The uncertainty from the size of the simulated samples is determined from pseudoexperiments
by varying the efficiency in each bin of each efficiency histogram within its uncertainties.
The signal and background components of the invariant mass fits are replaced with the
sum of two Gaussian functions and an exponential function, respectively,
to estimate systematic uncertainties due to the choice of models.

The largest  systematic uncertainties on $R_1$ and $R_2$ come from
the photon reconstruction (both 4.0\,\%)
and kinematic weighting of the simulated samples (2.2\,\% and 2.0\,\%, respectively).
For $R_{2/1}$, the largest systematic uncertainties are due to the size of the simulated samples
(1.2\,\%) and the treatment of multiple candidates (1.1\,\%).
The total systematic uncertainties on $R_1$, $R_2$ and $R_{2/1}$ are 5.2\,\%, 5.4\,\% and 2.0\,\%, respectively.

For the $\Lb$ mass measurement, systematic uncertainties are assigned due to
the uncertainty on the momentum scale for charged-particle tracks,
uncertainties on the $\chi_{cJ}$ and kaon masses,
energy loss in the material,
miscalibration of the electromagnetic calorimeter,
and the models for the signal and background components in the fit.
The total systematic uncertainty ($0.26\mevcc$) is dominated by the uncertainty on
the momentum scale ($0.24\mevcc$), the effect of which is determined by repeating the analysis
with the momentum scaling parameter varied up
and down by one standard deviation~\cite{LHCb-PAPER-2013-011}.
The uncertainty from the miscalibration of the electromagnetic calorimeter~\cite{Belyaev:2015ire} is found
to be small due to the mass constraints that are applied,
as is the uncertainty on the energy loss in material, which is taken from a previous
LHCb study~\cite{LHCb-PAPER-2015-060}.

In conclusion, the ratios of branching fractions are found to be
\begin{eqnarray*}
	R_1 &=& 0.242 \pm 0.014\pm0.013\pm0.009\,,\\
	R_2 &=& 0.248\pm 0.020\pm0.014 \pm 0.009\,,\\
	R_{2/1} &=& 1.02\phantom{0}\pm 0.10\phantom{0}\pm0.02\phantom{0}\pm0.05\,,
\end{eqnarray*}
where  the first uncertainty is statistical, the second systematic and the third
due to the uncertainty on the branching fractions of the $\chi_{cJ}\to\jpsi\g$ decays.
The values of $R_1$ and $R_2$ may be combined with existing measurements of
${\cal B}\left(\Lb\to\jpsi\proton\Km\right)/{\cal B}\left(\Bz\to\jpsi\Kstar(892)^0\right)$~\cite{LHCb-PAPER-2015-032} and
${\cal B}\left(\Bz\to\jpsi\Kstar(892)^0\right)$~\cite{Chilikin:2014bkk} to obtain absolute branching fraction measurements.
As the result in Ref.~\cite{Chilikin:2014bkk} assumes equal production of $\Bp\Bm$ and $\Bz\Bzb$ pairs
at the $\FourS$ resonance,
a correction is applied using the current world average value of ${\cal B}(\FourS\to\Bp\Bm)/{\cal B}(\FourS\to\Bz\Bzb) = 1.058 \pm 0.024$~\cite{PDG2016},
yielding ${\cal B}\left(\Bz\to\jpsi\Kstar(892)^0\right) = (1.22 \pm 0.08)\times10^{-3}$ and
${\cal B}\left(\Lb\to\jpsi\proton\Km\right) = (3.01\pm0.21\,^{+0.43}_{-0.26})\times10^{-4}$, where the second uncertainty is due to the ratio of fragmentation fractions, $f_{\Lb}/f_d$~\cite{LHCb-PAPER-2011-018,LHCb-PAPER-2014-004},
and the first incorporates all other sources. This gives
\begin{eqnarray*}
	{\cal B}\left(\Lb\to\chicone\proton\Km\right) &=& (7.3 \pm 0.4 \pm 0.4 \pm 0.6\,^{+1.0}_{-0.6})\times10^{-5}\,,\\
	{\cal B}\left(\Lb\to\chictwo\proton\Km\right) &=& (7.5 \pm 0.6 \pm 0.4 \pm 0.6\,^{+1.1}_{-0.6})\times10^{-5}\,,
\end{eqnarray*}
where the third uncertainty is due to uncertainties on the $\chi_{cJ}\to\jpsi\g$, $\Lb\to\jpsi\proton\Km$
and $\Bz\to\jpsi\Kstar(892)^0$ branching fractions and the fourth is due to $f_{\Lb}/f_d$~\cite{LHCb-PAPER-2011-018,LHCb-PAPER-2014-004}.
These results show no suppression of the \chictwo mode relative to the \chicone
mode in \Lb baryon decays, which is different to what is observed in
$\B\to\chi_{cJ}K$ decays~\cite{Mizuk:2008me,Aubert:2008ae,LHCb-PAPER-2013-024}.
These decays will be useful for future investigations into the nature of the two pentaquark candidates
observed by the LHCb collaboration and provide further information on the applicability of the
factorisation approach in describing $b$-hadron decays to final states containing charmonium.

The $\Lb$ mass has also been measured and is found to be $5619.44 \pm 0.28 \pm 0.26 \mevcc$,
where the first uncertainty is statistical and the second systematic. This result is
combined with previous \lhcb
measurements from $\Lb\to[c\overline{c}]X$ decays~\cite{LHCb-PAPER-2011-035,LHCb-PAPER-2012-048,LHCb-PAPER-2015-060}
assuming that systematic uncertainties on the momentum scale and
energy loss are fully correlated between the measurements while other sources of systematic uncertainties
are uncorrelated. The procedure for the combination is the same as that used in Ref.~\cite{LHCb-PAPER-2015-060}.
This yields a new average value of $5619.62 \pm 0.16 \pm 0.13 \mevcc$,
which supersedes previous combinations of these results.

\section*{Acknowledgements}

\noindent We express our gratitude to our colleagues in the CERN
accelerator departments for the excellent performance of the LHC. We
thank the technical and administrative staff at the LHCb
institutes. We acknowledge support from CERN and from the national
agencies: CAPES, CNPq, FAPERJ and FINEP (Brazil); MOST and NSFC (China);
CNRS/IN2P3 (France); BMBF, DFG and MPG (Germany); INFN (Italy);
NWO (The Netherlands); MNiSW and NCN (Poland); MEN/IFA (Romania);
MinES and FASO (Russia); MinECo (Spain); SNSF and SER (Switzerland);
NASU (Ukraine); STFC (United Kingdom); NSF (USA).
We acknowledge the computing resources that are provided by CERN, IN2P3 (France), KIT and DESY (Germany), INFN (Italy), SURF (The Netherlands), PIC (Spain), GridPP (United Kingdom), RRCKI and Yandex LLC (Russia), CSCS (Switzerland), IFIN-HH (Romania), CBPF (Brazil), PL-GRID (Poland) and OSC (USA). We are indebted to the communities behind the multiple open
source software packages on which we depend.
Individual groups or members have received support from AvH Foundation (Germany),
EPLANET, Marie Sk\l{}odowska-Curie Actions and ERC (European Union),
Conseil G\'{e}n\'{e}ral de Haute-Savoie, Labex ENIGMASS and OCEVU,
R\'{e}gion Auvergne (France), RFBR and Yandex LLC (Russia), GVA, XuntaGal and GENCAT (Spain), Herchel Smith Fund, The Royal Society, Royal Commission for the Exhibition of 1851 and the Leverhulme Trust (United Kingdom).

\addcontentsline{toc}{section}{References}
\ifx\mcitethebibliography\mciteundefinedmacro
\PackageError{LHCb.bst}{mciteplus.sty has not been loaded}
{This bibstyle requires the use of the mciteplus package.}\fi
\providecommand{\href}[2]{#2}

\newpage

\centerline{\large\bf LHCb collaboration}
\begin{flushleft}
\small
R.~Aaij$^{40}$,
B.~Adeva$^{39}$,
M.~Adinolfi$^{48}$,
Z.~Ajaltouni$^{5}$,
S.~Akar$^{59}$,
J.~Albrecht$^{10}$,
F.~Alessio$^{40}$,
M.~Alexander$^{53}$,
S.~Ali$^{43}$,
G.~Alkhazov$^{31}$
% ...

{\footnotesize \it
% ...
$ ^{5}$Clermont Universit{\'e}, Universit{\'e} Blaise Pascal, CNRS/IN2P3, LPC, Clermont-Ferrand, France\\
% ...
$ ^{10}$Fakult{\"a}t Physik, Technische Universit{\"a}t Dortmund, Dortmund, Germany\\
% ...
$ ^{31}$Petersburg Nuclear Physics Institute (PNPI), Gatchina, Russia\\
% ...
$ ^{39}$Universidad de Santiago de Compostela, Santiago de Compostela, Spain\\
$ ^{40}$European Organization for Nuclear Research (CERN), Geneva, Switzerland\\
% ...
$ ^{43}$Nikhef National Institute for Subatomic Physics, Amsterdam, The Netherlands\\
% ...
$ ^{48}$H.H. Wills Physics Laboratory, University of Bristol, Bristol, United Kingdom\\
% ...
$ ^{53}$School of Physics and Astronomy, University of Glasgow, Glasgow, United Kingdom\\
% ...
$ ^{59}$University of Cincinnati, Cincinnati, OH, United States\\
% ...
}
\end{flushleft}


\begin{mcitethebibliography}{10}
\mciteSetBstSublistMode{n}
\mciteSetBstMaxWidthForm{subitem}{\alph{mcitesubitemcount})}
\mciteSetBstSublistLabelBeginEnd{\mcitemaxwidthsubitemform\space}
{\relax}{\relax}

\bibitem{GellMann:1964nj}
M.~Gell-Mann, \ifthenelse{\boolean{articletitles}}{\emph{{A schematic model of
  baryons and mesons}},
  }{}\href{http://dx.doi.org/10.1016/S0031-9163(64)92001-3}{Phys.\ Lett.\
  \textbf{8} (1964) 214}\relax
\mciteBstWouldAddEndPuncttrue
\mciteSetBstMidEndSepPunct{\mcitedefaultmidpunct}
{\mcitedefaultendpunct}{\mcitedefaultseppunct}\relax
\EndOfBibitem
\bibitem{Zweig:1981pd}
G.~Zweig, \ifthenelse{\boolean{articletitles}}{\emph{{An SU(3) model for strong
  interaction symmetry and its breaking, Part 1}}, }{}\href{http://cds.\ cern.\
  ch/record/352337/}{CERN-TH-401} (1964)\relax
\mciteBstWouldAddEndPuncttrue
\mciteSetBstMidEndSepPunct{\mcitedefaultmidpunct}
{\mcitedefaultendpunct}{\mcitedefaultseppunct}\relax
\EndOfBibitem
\bibitem{Zweig:1964jf}
G.~Zweig, \ifthenelse{\boolean{articletitles}}{\emph{{An SU(3) model for strong
  interaction symmetry and its breaking, Part 2}}, }{}\href{https://cds.\
  cern.\ ch/record/570209}{CERN-TH-412}, reprinted in Developments in the Quark
  Theory of Hadrons{ \bf 1} (1980) 22 (ed.\ D.\ ~Lichtenberg and S.\
  ~Rosen)\relax
\mciteBstWouldAddEndPuncttrue
\mciteSetBstMidEndSepPunct{\mcitedefaultmidpunct}
{\mcitedefaultendpunct}{\mcitedefaultseppunct}\relax
\EndOfBibitem
\bibitem{LHCb-PAPER-2015-029}
LHCb collaboration, R.~Aaij {\em et~al.},
  \ifthenelse{\boolean{articletitles}}{\emph{{Observation of $\jpsi\proton$
  resonances consistent with pentaquark states in $\Lb\to\jpsi\proton\Km$
  decays}}, }{}\href{http://dx.doi.org/10.1103/PhysRevLett.115.072001}{Phys.\
  Rev.\ Lett.\  \textbf{115} (2015) 072001},
  \href{http://arxiv.org/abs/1507.03414}{{\normalfont\ttfamily
  arXiv:1507.03414}}\relax
\mciteBstWouldAddEndPuncttrue
\mciteSetBstMidEndSepPunct{\mcitedefaultmidpunct}
{\mcitedefaultendpunct}{\mcitedefaultseppunct}\relax
\EndOfBibitem
\bibitem{LHCb-PAPER-2016-009}
LHCb collaboration, R.~Aaij {\em et~al.},
  \ifthenelse{\boolean{articletitles}}{\emph{{Model-independent evidence for
  $\jpsi\proton$ contributions to $\Lb\to \jpsi\proton\Km$ decays}},
  }{}\href{http://dx.doi.org/10.1103/PhysRevLett.117.082002}{Phys.\ Rev.\
  Lett.\  \textbf{117} (2016) 082002},
  \href{http://arxiv.org/abs/1604.05708}{{\normalfont\ttfamily
  arXiv:1604.05708}}\relax
\mciteBstWouldAddEndPuncttrue
\mciteSetBstMidEndSepPunct{\mcitedefaultmidpunct}
{\mcitedefaultendpunct}{\mcitedefaultseppunct}\relax
\EndOfBibitem
\bibitem{Karliner:2015ina}
M.~Karliner and J.~L. Rosner, \ifthenelse{\boolean{articletitles}}{\emph{{New
  exotic meson and baryon resonances from doubly-heavy hadronic molecules}},
  }{}\href{http://dx.doi.org/10.1103/PhysRevLett.115.122001}{Phys.\ Rev.\
  Lett.\  \textbf{115} (2015) 122001},
  \href{http://arxiv.org/abs/1506.06386}{{\normalfont\ttfamily
  arXiv:1506.06386}}\relax
\mciteBstWouldAddEndPuncttrue
\mciteSetBstMidEndSepPunct{\mcitedefaultmidpunct}
{\mcitedefaultendpunct}{\mcitedefaultseppunct}\relax
\EndOfBibitem
\bibitem{Chen:2015moa}
H.-X. Chen {\em et~al.}, \ifthenelse{\boolean{articletitles}}{\emph{{Towards
  exotic hidden-charm pentaquarks in QCD}},
  }{}\href{http://dx.doi.org/10.1103/PhysRevLett.115.172001}{Phys.\ Rev.\
  Lett.\  \textbf{115} (2015) 172001},
  \href{http://arxiv.org/abs/1507.03717}{{\normalfont\ttfamily
  arXiv:1507.03717}}\relax
\mciteBstWouldAddEndPuncttrue
\mciteSetBstMidEndSepPunct{\mcitedefaultmidpunct}
{\mcitedefaultendpunct}{\mcitedefaultseppunct}\relax
\EndOfBibitem
\bibitem{Roca:2015dva}
L.~Roca, J.~Nieves, and E.~Oset,
  \ifthenelse{\boolean{articletitles}}{\emph{{LHCb pentaquark as a
  $\Dbar^*\Sigma_c-\Dbar^*\Sigma_c^*$ molecular state}},
  }{}\href{http://dx.doi.org/10.1103/PhysRevD.92.094003}{Phys.\ Rev.\
  \textbf{D92} (2015) 094003},
  \href{http://arxiv.org/abs/1507.04249}{{\normalfont\ttfamily
  arXiv:1507.04249}}\relax
\mciteBstWouldAddEndPuncttrue
\mciteSetBstMidEndSepPunct{\mcitedefaultmidpunct}
{\mcitedefaultendpunct}{\mcitedefaultseppunct}\relax
\EndOfBibitem
\bibitem{Maiani:2015vwa}
L.~Maiani, A.~D. Polosa, and V.~Riquer,
  \ifthenelse{\boolean{articletitles}}{\emph{{The new pentaquarks in the
  diquark model}},
  }{}\href{http://dx.doi.org/10.1016/j.physletb.2015.08.008}{Phys.\ Lett.\
  \textbf{B749} (2015) 289},
  \href{http://arxiv.org/abs/1507.04980}{{\normalfont\ttfamily
  arXiv:1507.04980}}\relax
\mciteBstWouldAddEndPuncttrue
\mciteSetBstMidEndSepPunct{\mcitedefaultmidpunct}
{\mcitedefaultendpunct}{\mcitedefaultseppunct}\relax
\EndOfBibitem
\bibitem{Lebed:2015tna}
R.~F. Lebed, \ifthenelse{\boolean{articletitles}}{\emph{{The pentaquark
  candidates in the dynamical diquark picture}},
  }{}\href{http://dx.doi.org/10.1016/j.physletb.2015.08.032}{Phys.\ Lett.\
  \textbf{B749} (2015) 454},
  \href{http://arxiv.org/abs/1507.05867}{{\normalfont\ttfamily
  arXiv:1507.05867}}\relax
\mciteBstWouldAddEndPuncttrue
\mciteSetBstMidEndSepPunct{\mcitedefaultmidpunct}
{\mcitedefaultendpunct}{\mcitedefaultseppunct}\relax
\EndOfBibitem
\bibitem{Li:2015gta}
G.-N. Li, X.-G. He, and M.~He, \ifthenelse{\boolean{articletitles}}{\emph{{Some
  predictions of diquark model for hidden charm pentaquark discovered at the
  LHCb}}, }{}\href{http://dx.doi.org/10.1007/JHEP12(2015)128}{JHEP \textbf{12}
  (2015) 128}, \href{http://arxiv.org/abs/1507.08252}{{\normalfont\ttfamily
  arXiv:1507.08252}}\relax
\mciteBstWouldAddEndPuncttrue
\mciteSetBstMidEndSepPunct{\mcitedefaultmidpunct}
{\mcitedefaultendpunct}{\mcitedefaultseppunct}\relax
\EndOfBibitem
\bibitem{Guo:2015umn}
F.-K. Guo, U.-G. Meissner, W.~Wang, and Z.~Yang,
  \ifthenelse{\boolean{articletitles}}{\emph{{How to reveal the exotic nature
  of the $P_c(4450)$}},
  }{}\href{http://dx.doi.org/10.1103/PhysRevD.92.071502}{Phys.\ Rev.\
  \textbf{D92} (2015) 071502},
  \href{http://arxiv.org/abs/1507.04950}{{\normalfont\ttfamily
  arXiv:1507.04950}}\relax
\mciteBstWouldAddEndPuncttrue
\mciteSetBstMidEndSepPunct{\mcitedefaultmidpunct}
{\mcitedefaultendpunct}{\mcitedefaultseppunct}\relax
\EndOfBibitem
\bibitem{Mikhasenko:2015vca}
M.~Mikhasenko, \ifthenelse{\boolean{articletitles}}{\emph{{A triangle
  singularity and the LHCb pentaquarks}},
  }{}\href{http://arxiv.org/abs/1507.06552}{{\normalfont\ttfamily
  arXiv:1507.06552}}\relax
\mciteBstWouldAddEndPuncttrue
\mciteSetBstMidEndSepPunct{\mcitedefaultmidpunct}
{\mcitedefaultendpunct}{\mcitedefaultseppunct}\relax
\EndOfBibitem
\bibitem{Liu:2015fea}
X.-H. Liu, Q.~Wang, and Q.~Zhao,
  \ifthenelse{\boolean{articletitles}}{\emph{{Understanding the newly observed
  heavy pentaquark candidates}},
  }{}\href{http://dx.doi.org/10.1016/j.physletb.2016.03.089}{Phys.\ Lett.\
  \textbf{B757} (2016) 231},
  \href{http://arxiv.org/abs/1507.05359}{{\normalfont\ttfamily
  arXiv:1507.05359}}\relax
\mciteBstWouldAddEndPuncttrue
\mciteSetBstMidEndSepPunct{\mcitedefaultmidpunct}
{\mcitedefaultendpunct}{\mcitedefaultseppunct}\relax
\EndOfBibitem
\bibitem{Meissner:2015mza}
U.-G. Meissner and J.~A. Oller,
  \ifthenelse{\boolean{articletitles}}{\emph{{Testing the $\chi_{c1}\, p$
  composite nature of the $P_c(4450)$}},
  }{}\href{http://dx.doi.org/10.1016/j.physletb.2015.10.015}{Phys.\ Lett.\
  \textbf{B751} (2015) 59},
  \href{http://arxiv.org/abs/1507.07478}{{\normalfont\ttfamily
  arXiv:1507.07478}}\relax
\mciteBstWouldAddEndPuncttrue
\mciteSetBstMidEndSepPunct{\mcitedefaultmidpunct}
{\mcitedefaultendpunct}{\mcitedefaultseppunct}\relax
\EndOfBibitem
\bibitem{Guo:2016bkl}
F.-K. Guo, U.~G. Meissner, J.~Nieves, and Z.~Yang,
  \ifthenelse{\boolean{articletitles}}{\emph{{Remarks on the $P_c$ structures
  and triangle singularities}},
  }{}\href{http://dx.doi.org/10.1140/epja/i2016-16318-4}{Eur.\ Phys.\ J.\
  \textbf{A52} (2016) 318},
  \href{http://arxiv.org/abs/1605.05113}{{\normalfont\ttfamily
  arXiv:1605.05113}}\relax
\mciteBstWouldAddEndPuncttrue
\mciteSetBstMidEndSepPunct{\mcitedefaultmidpunct}
{\mcitedefaultendpunct}{\mcitedefaultseppunct}\relax
\EndOfBibitem
\bibitem{Bayar:2016ftu}
M.~Bayar, F.~Aceti, F.-K. Guo, and E.~Oset,
  \ifthenelse{\boolean{articletitles}}{\emph{{A Discussion on Triangle
  Singularities in the $\Lambda_b \to J/\psi K^{-} p$ Reaction}},
  }{}\href{http://dx.doi.org/10.1103/PhysRevD.94.074039}{Phys.\ Rev.\
  \textbf{D94} (2016) 074039},
  \href{http://arxiv.org/abs/1609.04133}{{\normalfont\ttfamily
  arXiv:1609.04133}}\relax
\mciteBstWouldAddEndPuncttrue
\mciteSetBstMidEndSepPunct{\mcitedefaultmidpunct}
{\mcitedefaultendpunct}{\mcitedefaultseppunct}\relax
\EndOfBibitem
\bibitem{Mizuk:2008me}
Belle collaboration, R.~Mizuk {\em et~al.},
  \ifthenelse{\boolean{articletitles}}{\emph{{Observation of two resonance-like
  structures in the $\pi^+ \chi_{c1}$ mass distribution in exclusive $\Bdb \to
  \Km\pi^+ \chi_{c1}$ decays}},
  }{}\href{http://dx.doi.org/10.1103/PhysRevD.78.072004}{Phys.\ Rev.\
  \textbf{D78} (2008) 072004},
  \href{http://arxiv.org/abs/0806.4098}{{\normalfont\ttfamily
  arXiv:0806.4098}}\relax
\mciteBstWouldAddEndPuncttrue
\mciteSetBstMidEndSepPunct{\mcitedefaultmidpunct}
{\mcitedefaultendpunct}{\mcitedefaultseppunct}\relax
\EndOfBibitem
\bibitem{Aubert:2008ae}
BaBar collaboration, B.~Aubert {\em et~al.},
  \ifthenelse{\boolean{articletitles}}{\emph{{Evidence for $X(3872) \to
  \psi(2S) \gamma$ in $B^\pm \to X_{3872} K^\pm$ decays, and a study of $B \to
  c \bar{c} \gamma K$}},
  }{}\href{http://dx.doi.org/10.1103/PhysRevLett.102.132001}{Phys.\ Rev.\
  Lett.\  \textbf{102} (2009) 132001},
  \href{http://arxiv.org/abs/0809.0042}{{\normalfont\ttfamily
  arXiv:0809.0042}}\relax
\mciteBstWouldAddEndPuncttrue
\mciteSetBstMidEndSepPunct{\mcitedefaultmidpunct}
{\mcitedefaultendpunct}{\mcitedefaultseppunct}\relax
\EndOfBibitem
\bibitem{LHCb-PAPER-2013-024}
LHCb collaboration, R.~Aaij {\em et~al.},
  \ifthenelse{\boolean{articletitles}}{\emph{{Observation of
  $\Bs\to\chicone\phiz$ decay and study of $\Bz\to\chi_{c1,2}\Kstarz$ decays}},
  }{}\href{http://dx.doi.org/10.1016/j.nuclphysb.2013.06.005}{Nucl.\ Phys.\
  \textbf{B874} (2013) 663},
  \href{http://arxiv.org/abs/1305.6511}{{\normalfont\ttfamily
  arXiv:1305.6511}}\relax
\mciteBstWouldAddEndPuncttrue
\mciteSetBstMidEndSepPunct{\mcitedefaultmidpunct}
{\mcitedefaultendpunct}{\mcitedefaultseppunct}\relax
\EndOfBibitem
\bibitem{Beneke:2008pi}
M.~Beneke and L.~Vernazza, \ifthenelse{\boolean{articletitles}}{\emph{{$B \to
  \chi_{cJ} K$ decays revisited}},
  }{}\href{http://dx.doi.org/10.1016/j.nuclphysb.2008.11.025}{Nucl.\ Phys.\
  \textbf{B811} (2009) 155},
  \href{http://arxiv.org/abs/0810.3575}{{\normalfont\ttfamily
  arXiv:0810.3575}}\relax
\mciteBstWouldAddEndPuncttrue
\mciteSetBstMidEndSepPunct{\mcitedefaultmidpunct}
{\mcitedefaultendpunct}{\mcitedefaultseppunct}\relax
\EndOfBibitem
\bibitem{Bhardwaj:2015rju}
Belle collaboration, V.~Bhardwaj {\em et~al.},
  \ifthenelse{\boolean{articletitles}}{\emph{{Inclusive and exclusive
  measurements of $B$ decays to $\chi_{c1}$ and $\chi_{c2}$ at Belle}},
  }{}\href{http://dx.doi.org/10.1103/PhysRevD.93.052016}{Phys.\ Rev.\
  \textbf{D93} (2016) 052016},
  \href{http://arxiv.org/abs/1512.02672}{{\normalfont\ttfamily
  arXiv:1512.02672}}\relax
\mciteBstWouldAddEndPuncttrue
\mciteSetBstMidEndSepPunct{\mcitedefaultmidpunct}
{\mcitedefaultendpunct}{\mcitedefaultseppunct}\relax
\EndOfBibitem
\bibitem{Alves:2008zz}
LHCb collaboration, A.~A. Alves~Jr.\ {\em et~al.},
  \ifthenelse{\boolean{articletitles}}{\emph{{The \lhcb detector at the LHC}},
  }{}\href{http://dx.doi.org/10.1088/1748-0221/3/08/S08005}{JINST \textbf{3}
  (2008) S08005}\relax
\mciteBstWouldAddEndPuncttrue
\mciteSetBstMidEndSepPunct{\mcitedefaultmidpunct}
{\mcitedefaultendpunct}{\mcitedefaultseppunct}\relax
\EndOfBibitem
\bibitem{LHCb-DP-2014-002}
LHCb collaboration, R.~Aaij {\em et~al.},
  \ifthenelse{\boolean{articletitles}}{\emph{{LHCb detector performance}},
  }{}\href{http://dx.doi.org/10.1142/S0217751X15300227}{Int.\ J.\ Mod.\ Phys.\
  \textbf{A30} (2015) 1530022},
  \href{http://arxiv.org/abs/1412.6352}{{\normalfont\ttfamily
  arXiv:1412.6352}}\relax
\mciteBstWouldAddEndPuncttrue
\mciteSetBstMidEndSepPunct{\mcitedefaultmidpunct}
{\mcitedefaultendpunct}{\mcitedefaultseppunct}\relax
\EndOfBibitem
\bibitem{Sjostrand:2006za}
T.~Sj\"{o}strand, S.~Mrenna, and P.~Skands,
  \ifthenelse{\boolean{articletitles}}{\emph{{PYTHIA 6.4 physics and manual}},
  }{}\href{http://dx.doi.org/10.1088/1126-6708/2006/05/026}{JHEP \textbf{05}
  (2006) 026}, \href{http://arxiv.org/abs/hep-ph/0603175}{{\normalfont\ttfamily
  arXiv:hep-ph/0603175}}\relax
\mciteBstWouldAddEndPuncttrue
\mciteSetBstMidEndSepPunct{\mcitedefaultmidpunct}
{\mcitedefaultendpunct}{\mcitedefaultseppunct}\relax
\EndOfBibitem
\bibitem{Sjostrand:2007gs}
T.~Sj\"{o}strand, S.~Mrenna, and P.~Skands,
  \ifthenelse{\boolean{articletitles}}{\emph{{A brief introduction to PYTHIA
  8.1}}, }{}\href{http://dx.doi.org/10.1016/j.cpc.2008.01.036}{Comput.\ Phys.\
  Commun.\  \textbf{178} (2008) 852},
  \href{http://arxiv.org/abs/0710.3820}{{\normalfont\ttfamily
  arXiv:0710.3820}}\relax
\mciteBstWouldAddEndPuncttrue
\mciteSetBstMidEndSepPunct{\mcitedefaultmidpunct}
{\mcitedefaultendpunct}{\mcitedefaultseppunct}\relax
\EndOfBibitem
\bibitem{LHCb-PROC-2010-056}
I.~Belyaev {\em et~al.}, \ifthenelse{\boolean{articletitles}}{\emph{{Handling
  of the generation of primary events in Gauss, the LHCb simulation
  framework}}, }{}\href{http://dx.doi.org/10.1088/1742-6596/331/3/032047}{{J.\
  Phys.\ Conf.\ Ser.\ } \textbf{331} (2011) 032047}\relax
\mciteBstWouldAddEndPuncttrue
\mciteSetBstMidEndSepPunct{\mcitedefaultmidpunct}
{\mcitedefaultendpunct}{\mcitedefaultseppunct}\relax
\EndOfBibitem
\bibitem{Lange:2001uf}
D.~J. Lange, \ifthenelse{\boolean{articletitles}}{\emph{{The EvtGen particle
  decay simulation package}},
  }{}\href{http://dx.doi.org/10.1016/S0168-9002(01)00089-4}{Nucl.\ Instrum.\
  Meth.\  \textbf{A462} (2001) 152}\relax
\mciteBstWouldAddEndPuncttrue
\mciteSetBstMidEndSepPunct{\mcitedefaultmidpunct}
{\mcitedefaultendpunct}{\mcitedefaultseppunct}\relax
\EndOfBibitem
\bibitem{Golonka:2005pn}
P.~Golonka and Z.~Was, \ifthenelse{\boolean{articletitles}}{\emph{{PHOTOS Monte
  Carlo: A precision tool for QED corrections in $Z$ and $W$ decays}},
  }{}\href{http://dx.doi.org/10.1140/epjc/s2005-02396-4}{Eur.\ Phys.\ J.\
  \textbf{C45} (2006) 97},
  \href{http://arxiv.org/abs/hep-ph/0506026}{{\normalfont\ttfamily
  arXiv:hep-ph/0506026}}\relax
\mciteBstWouldAddEndPuncttrue
\mciteSetBstMidEndSepPunct{\mcitedefaultmidpunct}
{\mcitedefaultendpunct}{\mcitedefaultseppunct}\relax
\EndOfBibitem
\bibitem{Allison:2006ve}
Geant4 collaboration, J.~Allison {\em et~al.},
  \ifthenelse{\boolean{articletitles}}{\emph{{Geant4 developments and
  applications}}, }{}\href{http://dx.doi.org/10.1109/TNS.2006.869826}{IEEE
  Trans.\ Nucl.\ Sci.\  \textbf{53} (2006) 270}\relax
\mciteBstWouldAddEndPuncttrue
\mciteSetBstMidEndSepPunct{\mcitedefaultmidpunct}
{\mcitedefaultendpunct}{\mcitedefaultseppunct}\relax
\EndOfBibitem
\bibitem{Agostinelli:2002hh}
Geant4 collaboration, S.~Agostinelli {\em et~al.},
  \ifthenelse{\boolean{articletitles}}{\emph{{Geant4: A simulation toolkit}},
  }{}\href{http://dx.doi.org/10.1016/S0168-9002(03)01368-8}{Nucl.\ Instrum.\
  Meth.\  \textbf{A506} (2003) 250}\relax
\mciteBstWouldAddEndPuncttrue
\mciteSetBstMidEndSepPunct{\mcitedefaultmidpunct}
{\mcitedefaultendpunct}{\mcitedefaultseppunct}\relax
\EndOfBibitem
\bibitem{LHCb-PROC-2011-006}
M.~Clemencic {\em et~al.}, \ifthenelse{\boolean{articletitles}}{\emph{{The
  \lhcb simulation application, Gauss: Design, evolution and experience}},
  }{}\href{http://dx.doi.org/10.1088/1742-6596/331/3/032023}{{J.\ Phys.\ Conf.\
  Ser.\ } \textbf{331} (2011) 032023}\relax
\mciteBstWouldAddEndPuncttrue
\mciteSetBstMidEndSepPunct{\mcitedefaultmidpunct}
{\mcitedefaultendpunct}{\mcitedefaultseppunct}\relax
\EndOfBibitem
\bibitem{Breiman}
L.~Breiman, J.~H. Friedman, R.~A. Olshen, and C.~J. Stone, {\em Classification
  and regression trees}, Wadsworth international group, Belmont, California,
  USA, 1984\relax
\mciteBstWouldAddEndPuncttrue
\mciteSetBstMidEndSepPunct{\mcitedefaultmidpunct}
{\mcitedefaultendpunct}{\mcitedefaultseppunct}\relax
\EndOfBibitem
\bibitem{Hulsbergen:2005pu}
W.~D. Hulsbergen, \ifthenelse{\boolean{articletitles}}{\emph{{Decay chain
  fitting with a Kalman filter}},
  }{}\href{http://dx.doi.org/10.1016/j.nima.2005.06.078}{Nucl.\ Instrum.\
  Meth.\  \textbf{A552} (2005) 566},
  \href{http://arxiv.org/abs/physics/0503191}{{\normalfont\ttfamily
  arXiv:physics/0503191}}\relax
\mciteBstWouldAddEndPuncttrue
\mciteSetBstMidEndSepPunct{\mcitedefaultmidpunct}
{\mcitedefaultendpunct}{\mcitedefaultseppunct}\relax
\EndOfBibitem
\bibitem{PDG2016}
Particle Data Group, C.~Patrignani {\em et~al.},
  \ifthenelse{\boolean{articletitles}}{\emph{{\href{http://pdg.lbl.gov/}{Review
  of particle physics}}},
  }{}\href{http://dx.doi.org/10.1088/1674-1137/40/10/100001}{Chin.\ Phys.\
  \textbf{C40} (2016) 100001}\relax
\mciteBstWouldAddEndPuncttrue
\mciteSetBstMidEndSepPunct{\mcitedefaultmidpunct}
{\mcitedefaultendpunct}{\mcitedefaultseppunct}\relax
\EndOfBibitem
\bibitem{LHCb-DP-2012-003}
M.~Adinolfi {\em et~al.},
  \ifthenelse{\boolean{articletitles}}{\emph{{Performance of the \lhcb RICH
  detector at the LHC}},
  }{}\href{http://dx.doi.org/10.1140/epjc/s10052-013-2431-9}{Eur.\ Phys.\ J.\
  \textbf{C73} (2013) 2431},
  \href{http://arxiv.org/abs/1211.6759}{{\normalfont\ttfamily
  arXiv:1211.6759}}\relax
\mciteBstWouldAddEndPuncttrue
\mciteSetBstMidEndSepPunct{\mcitedefaultmidpunct}
{\mcitedefaultendpunct}{\mcitedefaultseppunct}\relax
\EndOfBibitem
\bibitem{Cowan:2016tnm}
G.~A. Cowan, D.~C. Craik, and M.~D. Needham,
  \ifthenelse{\boolean{articletitles}}{\emph{{RapidSim: An application for the
  fast simulation of heavy-quark hadron decays}},
  }{}\href{http://dx.doi.org/10.1016/j.cpc.2017.01.029}{Comput.\ Phys.\
  Commun.\  \textbf{214} (2017) 239},
  \href{http://arxiv.org/abs/1612.07489}{{\normalfont\ttfamily
  arXiv:1612.07489}}\relax
\mciteBstWouldAddEndPuncttrue
\mciteSetBstMidEndSepPunct{\mcitedefaultmidpunct}
{\mcitedefaultendpunct}{\mcitedefaultseppunct}\relax
\EndOfBibitem
\bibitem{Rogozhnikov:2016bdp}
A.~Rogozhnikov, \ifthenelse{\boolean{articletitles}}{\emph{{Reweighting with
  boosted decision trees}},
  }{}\href{http://dx.doi.org/10.1088/1742-6596/762/1/012036}{J.\ Phys.\ Conf.\
  Ser.\  \textbf{762} (2016) 012036},
  \href{http://arxiv.org/abs/1608.05806}{{\normalfont\ttfamily
  arXiv:1608.05806}}\relax
\mciteBstWouldAddEndPuncttrue
\mciteSetBstMidEndSepPunct{\mcitedefaultmidpunct}
{\mcitedefaultendpunct}{\mcitedefaultseppunct}\relax
\EndOfBibitem
\bibitem{Skwarnicki:1986xj}
T.~Skwarnicki, {\em {A study of the radiative cascade transitions between the
  Upsilon-prime and Upsilon resonances}}, PhD thesis, Institute of Nuclear
  Physics, Krakow, 1986,
  {\href{http://inspirehep.net/record/230779/}{DESY-F31-86-02}}\relax
\mciteBstWouldAddEndPuncttrue
\mciteSetBstMidEndSepPunct{\mcitedefaultmidpunct}
{\mcitedefaultendpunct}{\mcitedefaultseppunct}\relax
\EndOfBibitem
\bibitem{Wilks:1938dza}
S.~S. Wilks, \ifthenelse{\boolean{articletitles}}{\emph{{The large-sample
  distribution of the likelihood ratio for testing composite hypotheses}},
  }{}\href{http://dx.doi.org/10.1214/aoms/1177732360}{Ann.\ Math.\ Stat.\
  \textbf{9} (1938) 60}\relax
\mciteBstWouldAddEndPuncttrue
\mciteSetBstMidEndSepPunct{\mcitedefaultmidpunct}
{\mcitedefaultendpunct}{\mcitedefaultseppunct}\relax
\EndOfBibitem
\bibitem{Dalitz:1953cp}
R.~H. Dalitz, \ifthenelse{\boolean{articletitles}}{\emph{{On the analysis of
  $\tau$-meson data and the nature of the $\tau$-meson}},
  }{}\href{http://dx.doi.org/10.1080/14786441008520365}{Phil.\ Mag.\
  \textbf{44} (1953) 1068}\relax
\mciteBstWouldAddEndPuncttrue
\mciteSetBstMidEndSepPunct{\mcitedefaultmidpunct}
{\mcitedefaultendpunct}{\mcitedefaultseppunct}\relax
\EndOfBibitem
\bibitem{Govorkova:2015vqa}
E.~Govorkova, \ifthenelse{\boolean{articletitles}}{\emph{{Study of photon
  reconstruction efficiency using $B^+ \to J/\psi K^{(*)+}$ decays in the LHCb
  experiment}}, }{}\href{http://dx.doi.org/10.1134/S1063778816100070}{Phys.\
  Atom.\ Nucl.\  \textbf{79} (2016) 1474},
  \href{http://arxiv.org/abs/1505.02960}{{\normalfont\ttfamily
  arXiv:1505.02960}}\relax
\mciteBstWouldAddEndPuncttrue
\mciteSetBstMidEndSepPunct{\mcitedefaultmidpunct}
{\mcitedefaultendpunct}{\mcitedefaultseppunct}\relax
\EndOfBibitem
\bibitem{Pivk:2004ty}
M.~Pivk and F.~R. Le~Diberder,
  \ifthenelse{\boolean{articletitles}}{\emph{{sPlot: A statistical tool to
  unfold data distributions}},
  }{}\href{http://dx.doi.org/10.1016/j.nima.2005.08.106}{Nucl.\ Instrum.\
  Meth.\  \textbf{A555} (2005) 356},
  \href{http://arxiv.org/abs/physics/0402083}{{\normalfont\ttfamily
  arXiv:physics/0402083}}\relax
\mciteBstWouldAddEndPuncttrue
\mciteSetBstMidEndSepPunct{\mcitedefaultmidpunct}
{\mcitedefaultendpunct}{\mcitedefaultseppunct}\relax
\EndOfBibitem
\bibitem{LHCb-PAPER-2013-011}
LHCb collaboration, R.~Aaij {\em et~al.},
  \ifthenelse{\boolean{articletitles}}{\emph{{Precision measurement of $\D$
  meson mass differences}},
  }{}\href{http://dx.doi.org/10.1007/JHEP06(2013)065}{JHEP \textbf{06} (2013)
  065}, \href{http://arxiv.org/abs/1304.6865}{{\normalfont\ttfamily
  arXiv:1304.6865}}\relax
\mciteBstWouldAddEndPuncttrue
\mciteSetBstMidEndSepPunct{\mcitedefaultmidpunct}
{\mcitedefaultendpunct}{\mcitedefaultseppunct}\relax
\EndOfBibitem
\bibitem{LHCb-PAPER-2012-010}
LHCb collaboration, R.~Aaij {\em et~al.},
  \ifthenelse{\boolean{articletitles}}{\emph{{Measurement of relative branching
  fractions of $\B$ decays to $\psitwos$ and $\jpsi$ mesons}},
  }{}\href{http://dx.doi.org/10.1140/epjc/s10052-012-2118-7}{Eur.\ Phys.\ J.\
  \textbf{C72} (2012) 2118},
  \href{http://arxiv.org/abs/1205.0918}{{\normalfont\ttfamily
  arXiv:1205.0918}}\relax
\mciteBstWouldAddEndPuncttrue
\mciteSetBstMidEndSepPunct{\mcitedefaultmidpunct}
{\mcitedefaultendpunct}{\mcitedefaultseppunct}\relax
\EndOfBibitem
\bibitem{Belyaev:2015ire}
I.~M. Belyaev {\em et~al.},
  \ifthenelse{\boolean{articletitles}}{\emph{{Calibration of the LHCb
  electromagnetic calorimeter via reconstructing the neutral-pion invariant
  mass}}, }{}\href{http://dx.doi.org/10.1134/S1063778815090057}{Phys.\ Atom.\
  Nucl.\  \textbf{78} (2015) 1026}\relax
\mciteBstWouldAddEndPuncttrue
\mciteSetBstMidEndSepPunct{\mcitedefaultmidpunct}
{\mcitedefaultendpunct}{\mcitedefaultseppunct}\relax
\EndOfBibitem
\bibitem{LHCb-PAPER-2015-060}
LHCb collaboration, R.~Aaij {\em et~al.},
  \ifthenelse{\boolean{articletitles}}{\emph{{Observation of
  $\Lb\to\psitwos\proton\Km$ and $\Lb\to\jpsi\pip\pim\proton\Km$ decays and a
  measurement of the $\Lb$ baryon mass}},
  }{}\href{http://dx.doi.org/10.1007/JHEP05(2016)132}{JHEP \textbf{05} (2016)
  132}, \href{http://arxiv.org/abs/1603.06961}{{\normalfont\ttfamily
  arXiv:1603.06961}}\relax
\mciteBstWouldAddEndPuncttrue
\mciteSetBstMidEndSepPunct{\mcitedefaultmidpunct}
{\mcitedefaultendpunct}{\mcitedefaultseppunct}\relax
\EndOfBibitem
\bibitem{LHCb-PAPER-2015-032}
LHCb collaboration, R.~Aaij {\em et~al.},
  \ifthenelse{\boolean{articletitles}}{\emph{{Study of the productions of $\Lb$
  and $\Bzb$ hadrons in $\proton\proton$ collisions and first measurement of
  the $\Lb\to\jpsi\proton\Km$ branching fraction}},
  }{}\href{http://dx.doi.org/10.1088/1674-1137/40/1/011001}{Chin.\ Phys.\
  \textbf{C40} (2016) 011001},
  \href{http://arxiv.org/abs/1509.00292}{{\normalfont\ttfamily
  arXiv:1509.00292}}\relax
\mciteBstWouldAddEndPuncttrue
\mciteSetBstMidEndSepPunct{\mcitedefaultmidpunct}
{\mcitedefaultendpunct}{\mcitedefaultseppunct}\relax
\EndOfBibitem
\bibitem{Chilikin:2014bkk}
Belle collaboration, K.~Chilikin {\em et~al.},
  \ifthenelse{\boolean{articletitles}}{\emph{{Observation of a new charged
  charmoniumlike state in $\Bzb \to J/\psi K^-\pi^+$ decays}},
  }{}\href{http://dx.doi.org/10.1103/PhysRevD.90.112009}{Phys.\ Rev.\
  \textbf{D90} (2014) 112009},
  \href{http://arxiv.org/abs/1408.6457}{{\normalfont\ttfamily
  arXiv:1408.6457}}\relax
\mciteBstWouldAddEndPuncttrue
\mciteSetBstMidEndSepPunct{\mcitedefaultmidpunct}
{\mcitedefaultendpunct}{\mcitedefaultseppunct}\relax
\EndOfBibitem
\bibitem{LHCb-PAPER-2011-018}
LHCb collaboration, R.~Aaij {\em et~al.},
  \ifthenelse{\boolean{articletitles}}{\emph{{Measurement of $\bquark$ hadron
  production fractions in 7 TeV $\proton\proton$ collisions}},
  }{}\href{http://dx.doi.org/10.1103/PhysRevD.85.032008}{Phys.\ Rev.\
  \textbf{D85} (2012) 032008},
  \href{http://arxiv.org/abs/1111.2357}{{\normalfont\ttfamily
  arXiv:1111.2357}}\relax
\mciteBstWouldAddEndPuncttrue
\mciteSetBstMidEndSepPunct{\mcitedefaultmidpunct}
{\mcitedefaultendpunct}{\mcitedefaultseppunct}\relax
\EndOfBibitem
\bibitem{LHCb-PAPER-2014-004}
LHCb collaboration, R.~Aaij {\em et~al.},
  \ifthenelse{\boolean{articletitles}}{\emph{{Study of the kinematic
  dependences of $\Lb$ production in $\proton\proton$ collisions and a
  measurement of the $\Lb\to\Lc\pim$ branching fraction}},
  }{}\href{http://dx.doi.org/10.1007/JHEP08(2014)143}{JHEP \textbf{08} (2014)
  143}, \href{http://arxiv.org/abs/1405.6842}{{\normalfont\ttfamily
  arXiv:1405.6842}}\relax
\mciteBstWouldAddEndPuncttrue
\mciteSetBstMidEndSepPunct{\mcitedefaultmidpunct}
{\mcitedefaultendpunct}{\mcitedefaultseppunct}\relax
\EndOfBibitem
\bibitem{LHCb-PAPER-2011-035}
LHCb collaboration, R.~Aaij {\em et~al.},
  \ifthenelse{\boolean{articletitles}}{\emph{{Measurement of $\bquark$-hadron
  masses}}, }{}\href{http://dx.doi.org/10.1016/j.physletb.2012.01.058}{Phys.\
  Lett.\  \textbf{B708} (2012) 241},
  \href{http://arxiv.org/abs/1112.4896}{{\normalfont\ttfamily
  arXiv:1112.4896}}\relax
\mciteBstWouldAddEndPuncttrue
\mciteSetBstMidEndSepPunct{\mcitedefaultmidpunct}
{\mcitedefaultendpunct}{\mcitedefaultseppunct}\relax
\EndOfBibitem
\bibitem{LHCb-PAPER-2012-048}
LHCb collaboration, R.~Aaij {\em et~al.},
  \ifthenelse{\boolean{articletitles}}{\emph{{Measurements of the $\Lb$,
  $\Xibm$, and $\Omegab$ baryon masses}},
  }{}\href{http://dx.doi.org/10.1103/PhysRevLett.110.182001}{Phys.\ Rev.\
  Lett.\  \textbf{110} (2013) 182001},
  \href{http://arxiv.org/abs/1302.1072}{{\normalfont\ttfamily
  arXiv:1302.1072}}\relax
\mciteBstWouldAddEndPuncttrue
\mciteSetBstMidEndSepPunct{\mcitedefaultmidpunct}
{\mcitedefaultendpunct}{\mcitedefaultseppunct}\relax
\EndOfBibitem
\end{mcitethebibliography}
\end{document}